\documentclass[fleqn,usenatbib]{mnras}
\usepackage[english]{babel}
\usepackage[T1]{fontenc}
\usepackage{ae,aecompl}
\usepackage{fancyhdr}
\usepackage{amsfonts}
\usepackage{amsmath}
\usepackage{amssymb}
\usepackage{multicol}
\usepackage{layout}
\usepackage{graphicx}
\usepackage{multirow}
\usepackage{color}
\usepackage{multirow}
\usepackage{times}
\usepackage{natbib}
\def\be{\begin{equation}}
\def\ee{\end{equation}}

\usepackage[utf8]{inputenc}
\usepackage{lmodern}

\definecolor{valecol}{rgb}{0,0.5, 1.}

\newif\ifAMStwofonts
\AMStwofontstrue

\graphicspath{{./fig/}}

\title{Spherical Collapse Approach for Non-standard  Dark Matter Models and Enhanced Early Galaxy Formation in JWST }
\author[Davari, Ashoorioon, and Rezazadeh]{
	Zahra Davari$^{1,2}$, Amjad Ashoorioon$^{1}$, and  Kazem Rezazadeh$^{1}$\\ 
	$^1$School of Physics, Institute for Research in Fundamental Sciences (IPM),P.O. Box 19395-5531, Tehran, Iran\\
	$^2$School of Physics,Korea Institute for Advanced Study (KIAS),85 Hoegiro, Dongdaemun-gu,Seoul, 02455,Korea}
\begin{document}
	\label{firstpage}
	
	\maketitle	
	\begin{abstract}
	
	Using the spherical collapse approach, we investigate the impact of two alternative dark matter models, each characterized by distinct non-zero equations of state—one constant and the other time-dependent—on the nonlinear regime. Specifically, we compare these models to standard cold dark matter (CDM) by analyzing their influence on the linear density threshold for nonrelativistic component collapse and virial overdensity. Additionally, we explore the number count of collapsed objects, or dark matter halos, analogous to the number count of galaxy clusters. Finally, in light of recent discoveries by the James Webb Space Telescope (JWST), indicating the potential for more efficient early galaxy formation at higher redshifts, we investigate how alternative dark matter assumptions can enhance structure formation efficiency during the early universe.
	
	\end{abstract}
	
	
	
\section{Introduction and motivation}
\label{int}

The necessity for a non-baryonic matter component has been evident since the early 1930s, with observations such as the rotation curve of the galaxy \citep{Rubin:1970zza, Rubin:1980zd} and, on a cosmological scale, phenomena like gravitational lensing, large-scale structure (LSS), and the cosmic microwave background (CMB) \citep{Planck:2018nkj}. Despite extensive research in astrophysics and particle physics, the origin and nature of dark matter remain elusive.

Given the crucial role of dark matter in small scales and the nonlinear regime of gravitational clustering, exploring alternative models of cold dark matter and their distinct physical effects is advantageous \citep{Kopp:2016mhm, Tutusaus:2018noa, Pace:2019vrs,Davari:2022uwd,Arabameri:2023who}. In this context, our work investigates the nonlinear evolution of matter density perturbations within two alternative dynamical dark matter models. In 1972, Gunn and Gott introduced the spherical collapse model (SCM) as a semi-analytical and simple approach to studying the nonlinear evolution of overdensities in sub-horizon regions \citep{Gunn:1972sv}. By assuming spherically symmetric overdense regions with densities higher than the background dark matter density, the SCM offers insights into the formation of virialized halos. Initially, due to the small growth of perturbations, the linear theory of perturbations can describe the evolution of spherical overdensities properly. Due to self-gravity, we expect that the spherical overdense. However, as self-gravity causes the overdense region to expand more slowly than the Hubble flow, it becomes denser than the background.  At a certain redshift, known as the turnaround redshift, $z_{{}_{\rm ta}} $, the overdense region reaches a maximum radius then starts to compress afterward, and eventually collapses and completely decouples from the background expansion. After $z_{{}_{\rm ta}} $, the overdense spherical region collapses under its self-gravity to reach the steady state virial radius at a certain redshift $z_{{}_{\rm vir}}$~\citep{padmanabhan1993, Fosalba:1997tn}.

The evolution of spherical perturbations, from initial fluctuations to virialization, strongly depends on the background cosmology. Within the framework of standard General Relativity (GR), several studies have explored the SCM \citep{Basilakos:2009mz}, extending the analysis to scalar-tensor and modified gravities \citep{Fan:2015lta}. However, more complex numerical methods like N-body simulations, which require significant computational resources, are also used to study the nonlinear evolution of cosmic perturbations and the formation of large-scale structures \citep{Pace:2010sn}.

\citet{Pace:2019vrs} have discussed how changes in the  DM characteristic parameters, such as the equation of state $(w)$, and the effective sound speed $(c^2_{{}_s})$, affect the linear overdensity threshold for collapse $(\delta_{{}_c})$ and the virial overdensity $(\Delta_{{}_{\rm vir}})$. The study found that a relatively high value of the sound speed $(c^2_{{}_s}\sim 10^{-4})$ is required to significantly alter the evolution of $\delta_{{}_c}$ between general dark matter (GDM) and the standard cold dark matter (CDM) model. However, $w$ has a much stronger effect, especially on $\delta_{{}_c}$, for values around $w\sim 10^{-3}$. This is because $w$ modifies the background expansion history, while $c^2_{{}_s}$ primarily affects the perturbations. In this work, we study the nonlinear evolution of matter perturbations by focusing on two cold alternative dark matter models, one with a non-zero constant state equation similar to the model reviewed by~\citet{Pace:2019vrs} and the other with a dynamic equation of state and predicting the abundance of virialized halos in this model.

Recently, the JWST Cosmic Evolution Early Release Science (CEERS) program derived the cumulative stellar mass density at $7 <z < 10$ based on 14 galaxy candidates with masses in the range $\sim 10^9 -10^{10}M_\odot$ using the 1-5 $\mu$m coverage data~\citep{2023Natur.616..266L}. This release of early JWST observational data has challenged the current theory of galaxy formation under the $\Lambda$CDM model~\citep{Boylan-Kolchin:2022kae, Mason:2022tiy, Matsumoto:2022qri}. Specifically, \citet{Boylan-Kolchin:2022kae} have expressed this to be a serious tension with the standard galaxy formation theory. They recognized six galaxies with stellar masses $M_\star  = 10^{10 }\sim 10^{11}h^{-1}M_\odot$ at $7.4 \leq z \leq 9.1$. Their comoving cumulative stellar mass density is about 20 times higher at $z \sim 8$ and about three orders of magnitude higher at $z \sim 9$ than the prediction from the star formation theory in standard $\Lambda$CDM cosmology, which is based on previous observations. Although the huge abundance excess may be due to issues of galaxy selection, measurements of galaxy stellar mass and redshift, dust extinction, and sample variance, if future spectroscopic observation (for example through the follow-up observation by the JWST/NIRSpec) confirms this difference, it will be an important challenge to the $\Lambda$CDM model. Various solutions to solve this tension have been proposed so far, such as the rapidly accelerating primordial black holes~\citep{Hutsi:2022fzw,Yuan:2023bvh, Huang:2023chx}, the primordial non-Gaussianity in the initial conditions of cosmological perturbations~\citep{Biagetti:2022ode},  the Fuzzy Dark Matter~\citep{Gong:2022qjx}, the cosmic string~\citep{Jiao:2023wcn}, a blue-tilted primordial power spectrum~\citep{Parashari:2023cui},  and a gradual transition in the stellar initial mass function~\citep{2024MNRAS.529.3563T}. Our investigation seeks to address this tension by assuming alternative dark matter models.

This paper is structured as follows: in Sec.~\ref{sec:backeq}, we briefly introduce the equation of state of the considered dark matter models and describe the evolution of the Hubble flow in these models. In Sec.~\ref{sec:pereq}, we present the basic equations for the evolution of density perturbations in the linear and nonlinear regimes. In Sec.~\ref{sec:data},  we first investigate the linear and nonlinear density perturbations for the considered models to obtain the values of the cosmic parameters using the latest cosmological data. Then, in Sec.~\ref{sec:nonlin}, we study nonlinear structure formation. In Sec.~\ref{sec:halomass}, we compute the predicted mass function and the number count of halos using the Press-Schechter formalism. Subsequently, in Sec.~\ref{sec7}, we investigate the comoving stellar mass density contained within galaxies more massive than $M_\star$ and compare our findings with the JWST observations. Finally, in Sec.~\ref{concl}, we summarize our conclusions.


\section{Background dynamics}
\label{sec:backeq}

\begin{figure*}
\begin{center}
	\includegraphics[height=4cm,width=8cm]{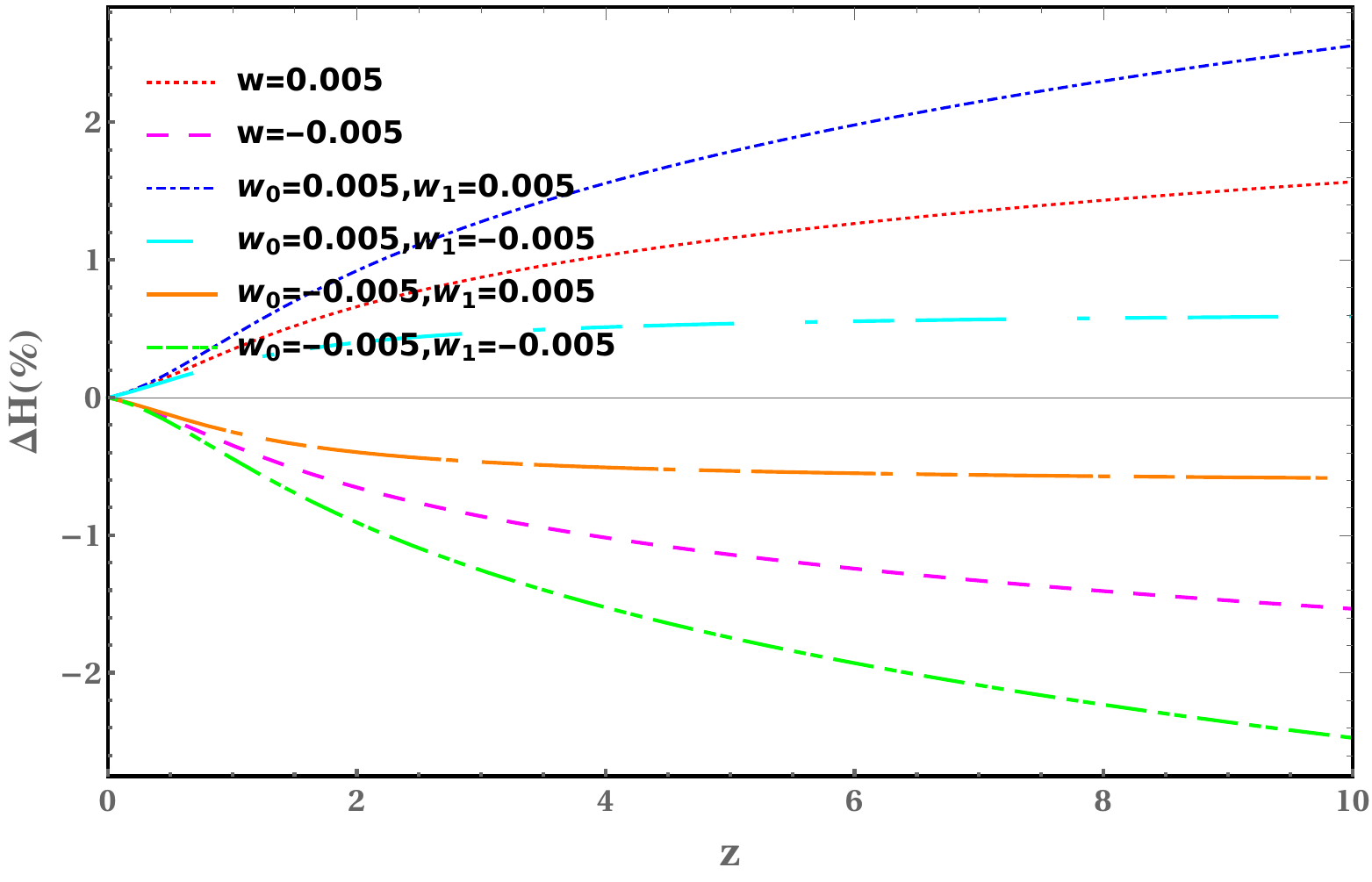}
	\includegraphics[height=4cm,width=8cm]{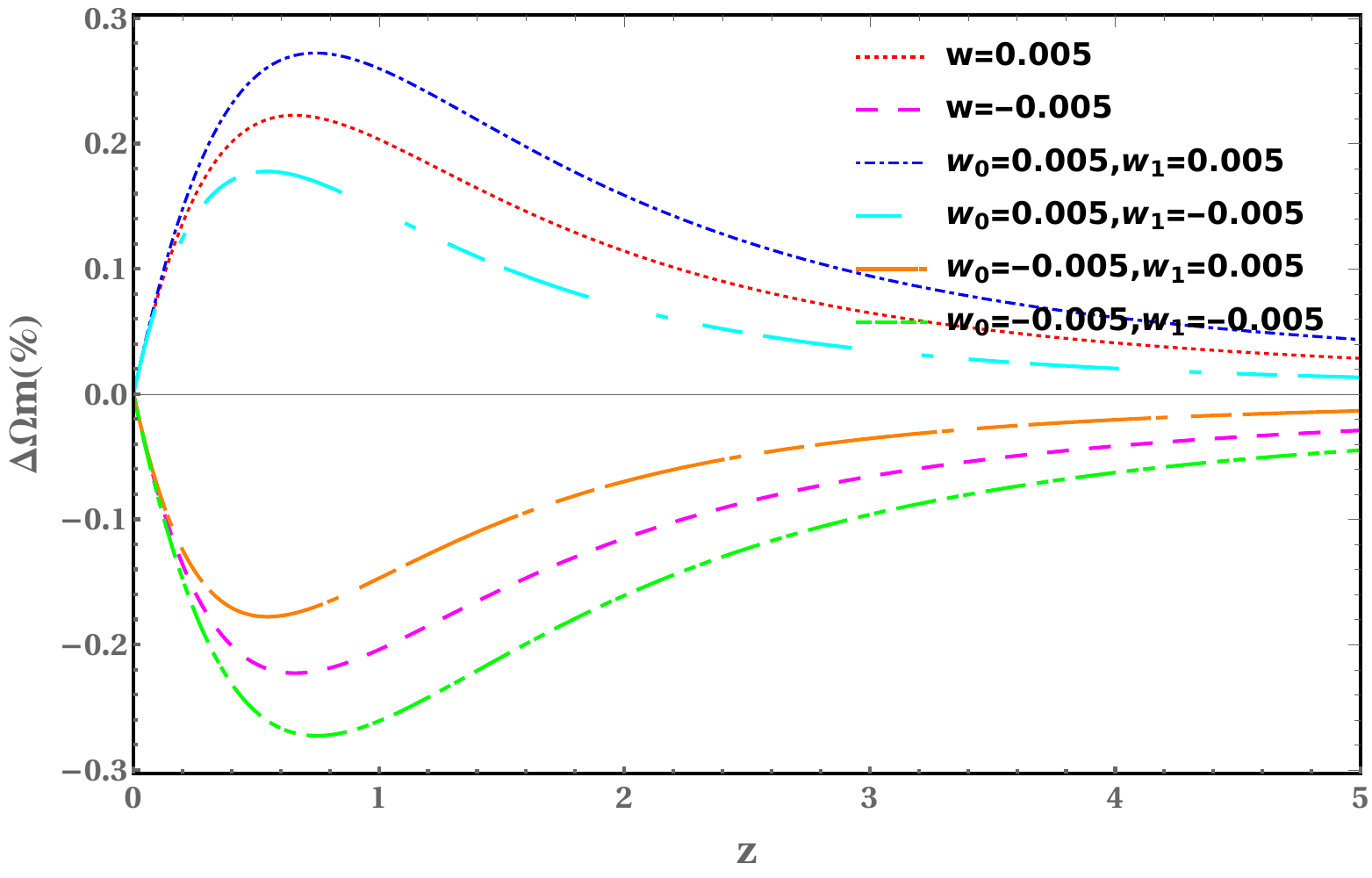}
	\caption{Relative deviation of the Hubble parameter (left panel) and the fractional energy density of dark matter (right panel) as a function of redshift, for various values of dark matter EoS   in  the proposed dark matter scenarios , compared to CDM model.}
	\label{fig1}
\end{center}
\end{figure*}

A flat, homogeneous, and isotropic Universe is described by the Friedmann-Lemaitre-Robertson-Walker (FLRW) metric
\begin{equation}\label{frw}
	ds^2=-dt^2+a^2(t)(dr^2+r^2d\theta^2+r^2\sin^2\theta d\phi^2) \, ,
\end{equation}
where $a(t)$ is the scale factor at the cosmic time $t$. In the framework of general relativity, we can obtain the Friedmann background dynamics equations by inserting equation~\eqref{frw} into the Einstein field equations, yielding:
\begin{align}
	&H^{2}=\frac{8\pi G}{3}\,\Sigma\rho_{{}_{j}} \, ,\\
	&\dot{H}=-4\pi G\,\Sigma(\rho_{{}_{j}}+p_{{}_{j}}) \, ,
\end{align}
where $H\equiv\dot{a}/a$ is the Hubble parameter, and $\rho_{{}_{j}}$ and $ p_{{}_{j}}$, respectively,  represent the mean energy density and mean pressure of each component $j$, respectively.  We assume that the material content of the Universe is composed of four components: radiation, baryons, dark matter, and dark energy, all described by ideal fluids with the equation of state $w _{{}_j}=p_{{}_j} /\rho_{{}_j}$. The energy densities of these components satisfy the following conservation equation:
\begin{equation}
	\label{rhodoti}
	\dot{\rho}_{{}_j}+3H(\rho_{{}_j}+p_{{}_j})=0 \, .
\end{equation}
The radiation component (photons($\gamma$)+relativistic neutrinos) will be denoted by the sub-index "R" and is characterized by a state parameter $w_{{}_R}=1/3$. We set $\Omega_{{}_{R,0}}=\Omega_{{}_{\gamma,0}}(1 + 0.2271 N_{{}_{\rm eff}})$, where $N_{\rm eff}$ is the effective extra relativistic degrees of freedom, taken as 3.046 in agreement with the standard model prediction~\citep{Planck:2018vyg}. We further take $\Omega_{{}_{\gamma,0}}=2.469\times10^{-5}h^{-2}$, where the reduced Hubble constant $h$ is defined as usual according to $H_{{}_{0}}\equiv100h\,\mathrm{km\,s^{-1}Mpc^{-1}}$. The baryon component, denoted by the sub-index "B", is a pressureless component, so $w_{{}_B}=0$. The dark energy component is considered as a cosmological constant, $\Lambda$, with a constant equation of state (EoS) parameter $w_{{}_\Lambda}=-1$. Lastly, the dark matter component is denoted by the sub-index "DM". It is considered in this study  in the following ways:
\begin{itemize}
	\item For the first model, we consider dark matter as a non-luminous, dark component of matter that must be non-relativistic or cold, similar to the standard model of cosmology $\Lambda$CDM. This means that cold dark matter is pressureless matter fluid, so $w_{{}_{DM}}=0$. For a flat Universe through the Friedmann equation,  the Hubble parameter be written as
	\begin{equation}
		H^2(a)=H_{{}_{0}}^2\left(\Omega_{{}_{R,0}} a^{-4}+\Omega_{{}_{B,0}} a^{-3}+\Omega_{{}_{DM,0}} a^{-3}+\Omega_\Lambda\right) \, .
	\end{equation}
	\item For the dynamical dark matter model, we consider the simplest model, that dark matter has a constant non-zero equation of state. Using the continuity equation  $\dot{\rho}=-3H(1+w)\rho$, the Hubble parameter can be written as:
	\begin{equation}
		H^2(a)=H_{{}_0}^2\left(\Omega_{{}_{R,0}} a^{-4}+\Omega_{{}_{B,0}} a^{-3}+\Omega_{{}_{DM,0}} a^{-3(1+w)}+\Omega_{{}_\Lambda}\right) \, .
	\end{equation}
	\item To explore whether the dark matter equation of state evolves, a simple approach is to consider the dark matter EoS as a function of redshift. Specifically, we consider the so-called Chevallier-Polarski-Linder (CPL) parametrization $w(a)=w_{{}_0}+w_{{}_1}(1-a)$, where the parameter $ w_{{}_1} $ describes the rate of evolution of dark matter~\citep{Chevallier:2000qy}.  The Hubble parameter is then
	\begin{align}
		H^2(a) = & H_{_{0}}^{2}\Big(\Omega_{_{R,0}}a^{-4}+\Omega_{_{B,0}}a^{-3} \nonumber \\
	& +\Omega_{_{DM,0}}a^{-3(1+w_{0}+w_{1})}e^{3w_{1}(a-1)}+\Omega_{_{\Lambda}}\Big) \, .
	\end{align}
This case reduces to cold dark matter when $w_{{}_0}=w_{{}_1}=0$.
\end{itemize}
In all the models, due to the flatness of the Universe, we have
\begin{equation}
	\label{OmegaLambda}
	\Omega_{{}_\Lambda}=1-\Omega_{{}_{R,0}}-\Omega_{{}_{B,0}}-\Omega_{{}_{DM,0}}  \, .
\end{equation}
To gain a clearer understanding of the impact of a non-zero dark matter equation of state compared to the cold dark matter model, we have plotted the evolution of two key cosmic quantities -the Hubble parameter and the dimensionless matter density- for various values of dark matter EoS  in Figure \ref{fig1}. As shown in the right panel of Figure \ref{fig1}, the quantity $\Delta H = \left[\frac{H_{NDM,i} - H_{CDM}}{H_{CDM}}\right] \times 100$ is positive (negative) for models with $w(a=1) > 0$ ($w(a=1) < 0$) across all redshifts. This indicates that the cosmic expansion is larger (smaller) compared to the concordance CDM model. A similar trend is observed for $\Delta \Omega_m$ in the right panel of Figure \ref{fig1}, with a noticeable deviation around $z \sim 0.7$. This deviation reaches approximately $\pm 0.3$ for the parameter sets $[w_0 = +0.005, w_1 = +0.005]$ and $[w_0 = -0.005, w_1 = -0.005]$.

\section{Cosmological perturbations}
\label{sec:pereq}

The large-scale structures in the Universe provide invaluable information about the nature of unknown dark components, in addition to the background evolution. The primordial matter perturbations grow throughout cosmic history, and their growth rate depends on the overall energy budget and the properties of the cosmic fluids.

By assuming the stress-energy tensor of a  perfect fluid for each component $j$ as
\begin{equation}
	T_j^{\mu\nu} = (\rho_j+P_j)u^{\mu} u^{\nu}+P_j g^{\mu\nu} \, ,
\end{equation}
where $u^\mu$ is the four-velocity of each fluid. We define the density contrast between a single fluid by the relation $\delta_{j}+1=\rho_{j}/\bar{\rho}$ with the background density $\bar{\rho}$ and the three-dimensional comoving peculiar velocity $\vec{u}$. Contracting the continuity equation, $\bigtriangledown_\mu T^{\mu\nu} = 0$, once with  $\vec{u}$ and once with the projection operator $h_{\mu\alpha} = g_{\mu\alpha}+u_\mu u_\alpha$, we can  derive the continuity and the Euler equations (the perturbation equations) as follows:
\begin{align}
	&\dot{\delta}_{j}+3H(c_{s,j}^2-w_j)\delta_{ j}+\left[1+w_j+(1+c_{s,j}^2)\delta_j\right]\vec{\nabla}.\vec{u}_j=0 \, \label{eqdelta},\\
	&\dot{\vec{u}}_{ j}+2H\vec{u}_j+(\vec{u}_j.\vec{\nabla})\vec{u}_j+\frac{\vec{\nabla}\phi}{a^2}=0  \, ,\label{theta}
\end{align}
where  $c_{s}^2=\delta p/\delta \rho$ is the effective sound speed and the dot denotes the derivative with respect to the physical time\citep{Ma:1995ey, Abramo:2008ip}. The peculiar gravitational potential is denoted by $\phi$, and it satisfies the Poisson equation obtained from the 00-component of Einstein's field equations
\begin{equation}
	\nabla^2\phi=4\pi G\Sigma  \rho_j(1+3c^2_{s,j})\delta_j \, .
\end{equation}
The divergence of equation~\eqref{theta} can be
written as
\begin{equation}
	\dot{\theta}_j+2H\theta_j + \frac{1}{3}\theta_j^2+\frac{\nabla^2\phi}{a^2} = 0 \, ,
\end{equation}
where $\theta_j\equiv\vec{\nabla}.\vec{u}_j$ and we have used $ \vec{\nabla}.[(\vec{u}_j.\vec{\nabla})\vec{u}_j] = \frac{1}{3}\theta_i^2$.  We can also rewrite equation~\eqref{eqdelta} as
\begin{equation}
	\dot{\delta}_j+ 3H(c^2_{s,j} - w_j)\delta_i + [1 + w_j+(1 + c^2_{ s,j})\delta_j]\theta_j = 0 \,.
\end{equation}
As mentioned in the previous section, we assumed dark matter as a non-zero pressure fluid, with the EoS parameter $w$. Since we do not have an evolution equation for $\delta p$, we approximate $c_{\rm s}^2$ by the adiabatic sound speed $c^2_{{\rm ad}}=w-\frac{\dot{w}}{3H(1+w)}$. Replacing the equation \eqref{theta} with the equation \eqref{eqdelta}, and eliminating $\theta_j$ from the system of equations and introducing quantities
\begin{equation}
	A_j \equiv3H(c^2_{s,j} - w_j),
	\qquad B_j \equiv \delta_j + [1 + w_j+(1 + c^2_{s,j})\delta_j]\theta_j \, ,
	\nonumber
\end{equation}
we can obtain the following equation:
\begin{equation}\label{pertot}
	\ddot{\delta}_j+\left(2H-\frac{\dot{B}_j}{B_j}\right)(A_j+\dot{\delta}_j)- \frac{1}{3}\frac{(\dot{\delta}_j+A_j)^2}{B_j}-\frac{B_j}{a^2}\nabla^2\phi = 0 \, .
\end{equation}
Since we consider dark energy as the cosmological constant and it does not change in space and time, it cannot cluster like matter, so $\delta_{\Lambda}=0$. Using the relation $\frac{d}{dt}=aH\frac{d}{da}$, equation~\eqref{pertot} can be written in terms of the scale factor as
\begin{eqnarray}\label{perm}
	&&\delta''_m+ A_m\delta'_m+B_m\delta_m-\frac{3}{2a^2}\Omega_m(a)  
	\nonumber \\
	&&\left[1+w+(1+c^2_s)\delta_m\right](1+3c^2_s)\delta_m(1+\delta_m)
	\nonumber \\
	&&=C_m\delta^2_m+D_m\frac{\delta'^2_m}{1+\delta_m} \, ,
\end{eqnarray}
where prime denotes derivative with respect to $a$ and the coefficients are
\begin{align}
	&A_m = \frac{3}{a}(1+c^2_s - w)+\frac{H'}{H}-\frac{w'}{1+w+(1+c^2_s)\delta_m} \, ,\nonumber\\
	&B_m = \frac{3}{a}\left[(c^2_s-w)(\frac{2}{a}+\frac{H'}{H})-w'-\frac{w'(c^2_s-w)}{1+w+(1+c^2_s)\delta_m}\right] \, ,\nonumber\\
	&C_m = \frac{3}{a^2}\frac{c^2_s-w}{1+w+(1+c^2_s)\delta_m},\nonumber\\
	&D_m = \frac{4+3c^2_s}{3(1+w+(1+c^2_s)\delta_m)} \, .\nonumber
\end{align}
As expected, in the case of cold dark matter model
($w = c^2_s = 0$), equation~\eqref{perm}  reduces to the well-known differential equation ( see \cite{Abramo:2008ip})
\begin{equation}
	\delta''_m+ \left(\frac{3}{a}+\frac{H'}{H}\right)\delta'_m-\frac{3}{2a^2}\Omega_m(a)\delta_m(1+\delta_m)=\frac{4}{3}\frac{\delta'^2_m}{1+\delta_m} \, . 	
\end{equation}
In the following sections, we will solve this differential equation to find the evolution of the matter density contrast.


\section{Alternative models of dark matter versus the cosmological data}
\label{sec:data}

To check the spherical collapse model, we need the constrained values of the free parameters of the model and the cosmological parameters. Therefore, we aim to confine the models introduced in the previous section to the background and perturbation levels using the latest cosmological data via the Markov Chain Monte Carlo (MCMC) method.  For this purpose, we implement the related equations in the publicly available numerical code \texttt{CLASS}\footnote{\label{myfootnote0}\url{https://github.com/lesgourg/class_public}} (Cosmic Linear Anisotropy Solving System)~\citep{Lesgourgues:2011rh} and using the code \texttt{MONTEPYTHON-v3}\footnote{\label{myfootnote}\url{https://github.com/baudren/montepython_public}}~\citep{Audren:2012wb, Brinckmann:2018cvx} to perform a Monte Carlo Markov chain analysis with a Metropolis-Hasting algorithm. In our analysis, we use the combinations of data points including luminosity distance of type Ia supernovas (SNIa, 1048 points from the Pantheon sample)~\citep{Pan-STARRS1:2017jku}, cosmic chronometers ($H(z)$, 30 points from Table 4 of~\citep{Moresco:2016mzx}), measurements of the growth rate of structure weighted by a redshift-dependent normalization, $ f(z) \sigma_8(z) $, obtained by fitting WiggleZ Survey data and  SDSS-III Baryon Oscillation Spectroscopic Survey~\citep{BOSS:2016wmc}, and the CMB temperature and polarization angular power spectra (the high-l TT, TE, EE +low-l TT, EE+lensing data from Planck 2018)~\citep{Planck:2018vyg}.

We report the result of our MCMC analysis for the three models in Tables~\ref{best} and \ref{tab:comp}. As mentioned in the section~\ref{int}, in \citep{Kopp:2016mhm, Tutusaus:2018noa,Pace:2019vrs} is replaced CDM with the Generalized Dark Matter	(GDM) model to introduce three parametric functions to the model: the GDM equation of state $w$, the sound speed $c^2_s$ and the viscosity $c^2_{vis}$. The equation of state $w$ is reported  $-0.000896 < w < 0.00238$ at the 99.7\% CL,  $0.074^{+0.111}_{-0.110}(10^{-2})$, and $0.055\pm 0.053(10^{-2})$ at the 95\% CL and in \citep{Kopp:2016mhm},\citep{Tutusaus:2018noa}, and \citep{Pace:2019vrs}, respectively . We obtained the value of $w$ for the NDM-cte model consistent with these results.

\begin{figure*}
\begin{center}
	\includegraphics[height=7cm,width=8cm]{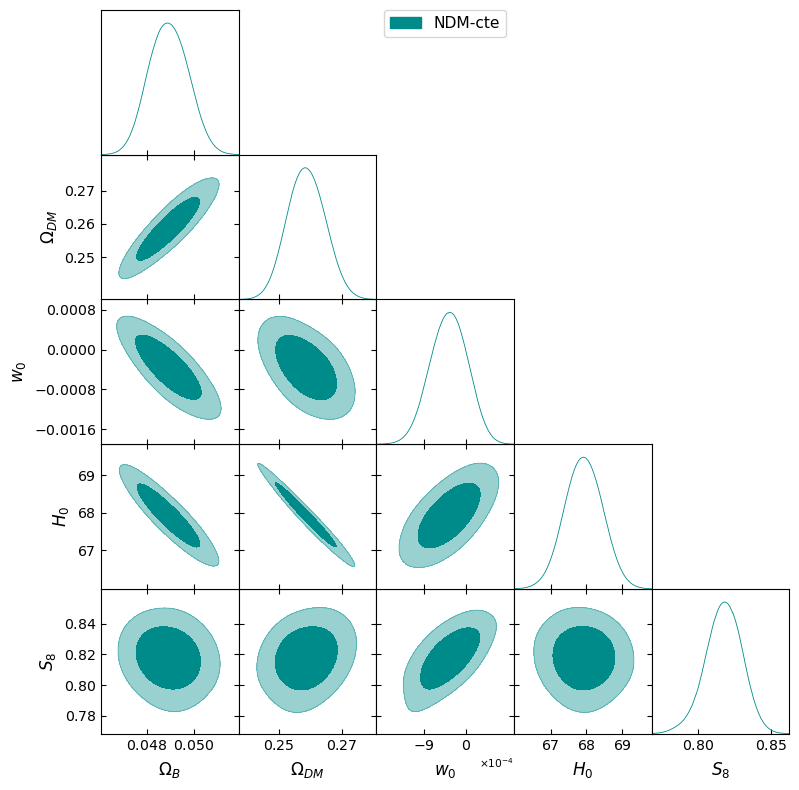}
	\includegraphics[height=7cm,width=8cm]{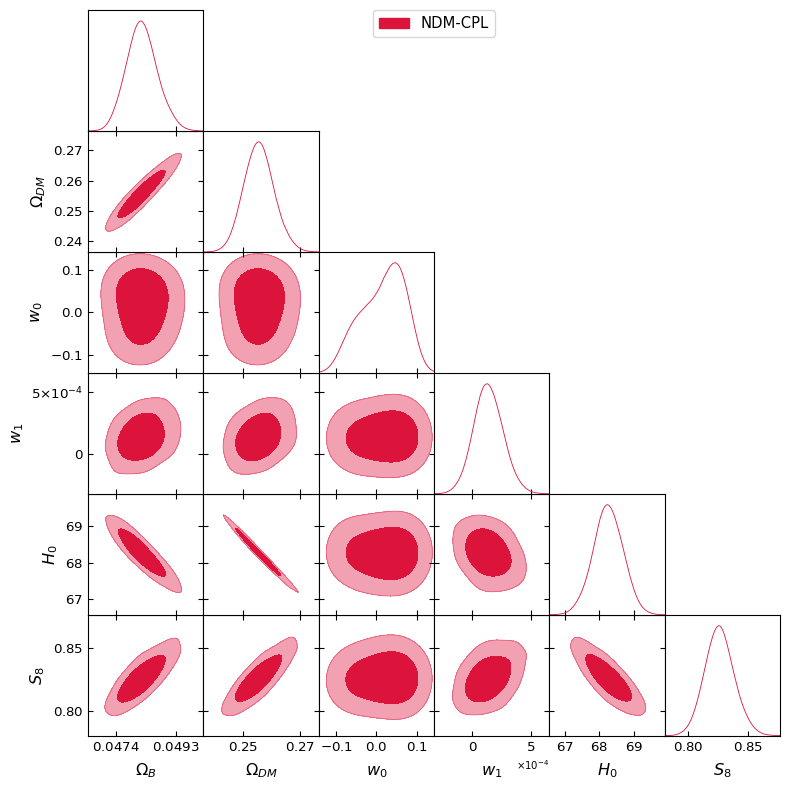}
	\includegraphics[width=8cm]{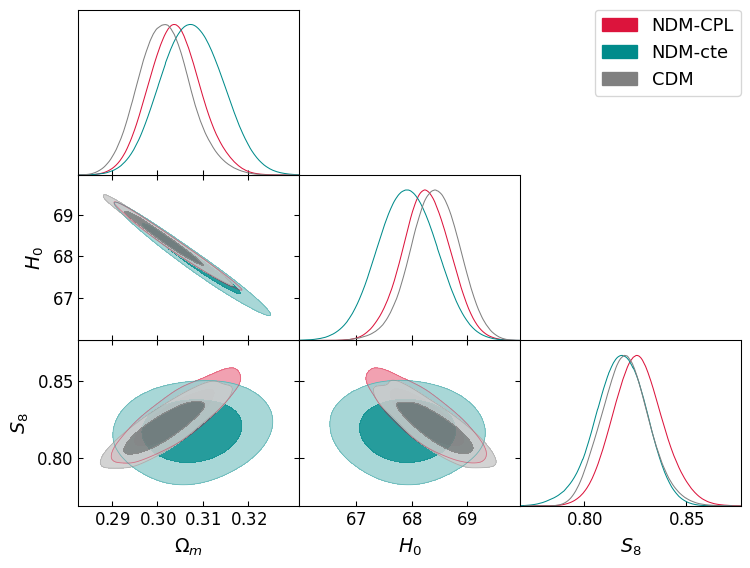}
	\caption{Reconstructed 2D posterior distribution in the considered model.}
	\label{figndm-cpl}
\end{center}
\end{figure*}

Additionally, we show 1D posterior distributions and 2D contours of the free parameters for all the models in Figure~\ref{figndm-cpl}. Given the Jeffreys' scale~\citep{Rivera:2016zzr}, since the difference between the AIC of a given model and the CDM model is smaller than 4, we conclude that the models examined in this work and the CDM model are equally supported by the observational data.
\begin{table*}
	\centering
	\caption{The best values of the free parameters obtained for the considered models.}
	\begin{tabular}{|l|c|c|c|c|}
		\hline
		NDM-cte & best-fit & mean$\pm\sigma$ & 95\% lower & 95\% upper \\ \hline
		$\Omega_\mathrm{B}$ &$0.04891$ & $0.04906_{-0.00079}^{+0.00079}$ & $0.04701$ & $0.5092$ \\
		$\Omega_\mathrm{DM}$ &$0.2586$ & $0.2596_{-0.0046}^{+0.0069}$ & $0.2473$ & $0.2689$ \\
		$H_0$ &$67.93$ & $67.84_{-0.54}^{+0.42}$ & $66.93$ & $68.93$ \\
		$S_8$ &$0.817$ & $0.818\pm0.012$ & $0.790$ & $0.842$ \\
		$w_0$ &$-0.0003626$ & $-0.0004322_{-0.00043}^{+0.00033}$ & $-0.001285$ & $0.0005327$\\ \hline
	NDM-CPL & best-fit & mean$\pm\sigma$ & 95\% lower & 95\% upper \\ \hline
	$\Omega{}_{B }$ &$0.0481$ & $0.04826_{-0.00058}^{+0.00047}$ & $0.0473$ & $0.04933$ \\
	$\Omega{}_{DM }$ &$0.2536$ & $0.256_{-0.0061}^{+0.0049}$ & $0.2454$ & $0.2672$ \\
	$H_0$ &$68.39$ & $68.23_{-0.4}^{+0.49}$ & $67.32$ & $69.09$ \\
	$S_8$ &$0.822$ & $0.826\pm0.012$ & $0.803$ & $0.851$ \\
	$10^{+4}w_0$ &$0.01738$ & $0.04403^{+0.068}_{-0.046}$ & $-0.113$ & $0.127$ \\
	$10^{+4}w_1$ &$0.0001436$ & $0.0001399_{-0.00013}^{+0.00012}$ & $-0.0001166$ & $0.0004044$ \\
	\hline
	CDM & best-fit & mean$\pm\sigma$ & 95\% lower & 95\% upper \\ \hline
	$\Omega{}_{B }$ &$0.04836$ & $0.04806_{-0.00055}^{+0.00047}$ & $0.04707$ & $0.04912$ \\
	$\Omega{}_{DM }$ &$0.2561$ & $0.2536_{-0.0062}^{+0.0049}$ & $0.2427$ & $0.2653$ \\
	$H_0$ &$68.24$ & $68.42_{-0.44}^{+0.5}$ & $67.48$ & $69.3$ \\
	$S_8$ &$0.826$ & $0.820\pm0.012$ & $0.797$ & $0.843$ \\
	\hline
\end{tabular}\label{best}
\end{table*}
\begin{table}
\centering
\caption{The result of MCMC analysis for all the reviewed models in this study for the best-fit $\chi^2$.}
\begin{tabular}{|llll|}                                   \hline                           		\multicolumn{1}{|l|}{}                           & \multicolumn{1}{l|}{CDM} & \multicolumn{1}{l|}{NDM-cte} & \multicolumn{1}{l|}{NDM-CPL} \\ \hline
	\multicolumn{1}{|l|}{$\chi^2_{\rm min}$}         & \multicolumn{1}{l|}{$ 4639.34$}   & \multicolumn{1}{l|}{$ 4638.00 $}   & \multicolumn{1}{l|}{$ 4637.26 $}     \\ \hline
	\multicolumn{1}{|l|}{$\rm AIC-AIC_{\rm CDM} $}                           & \multicolumn{1}{l|}{$-$}   & \multicolumn{1}{l|}{$ 0.66 $}   & \multicolumn{1}{l|}{$ 1.92 $}      \\ \hline
\end{tabular}\label{tab:comp}
\end{table}
\begin{figure*}
\begin{center}
	\includegraphics[height=4cm,width=8cm]{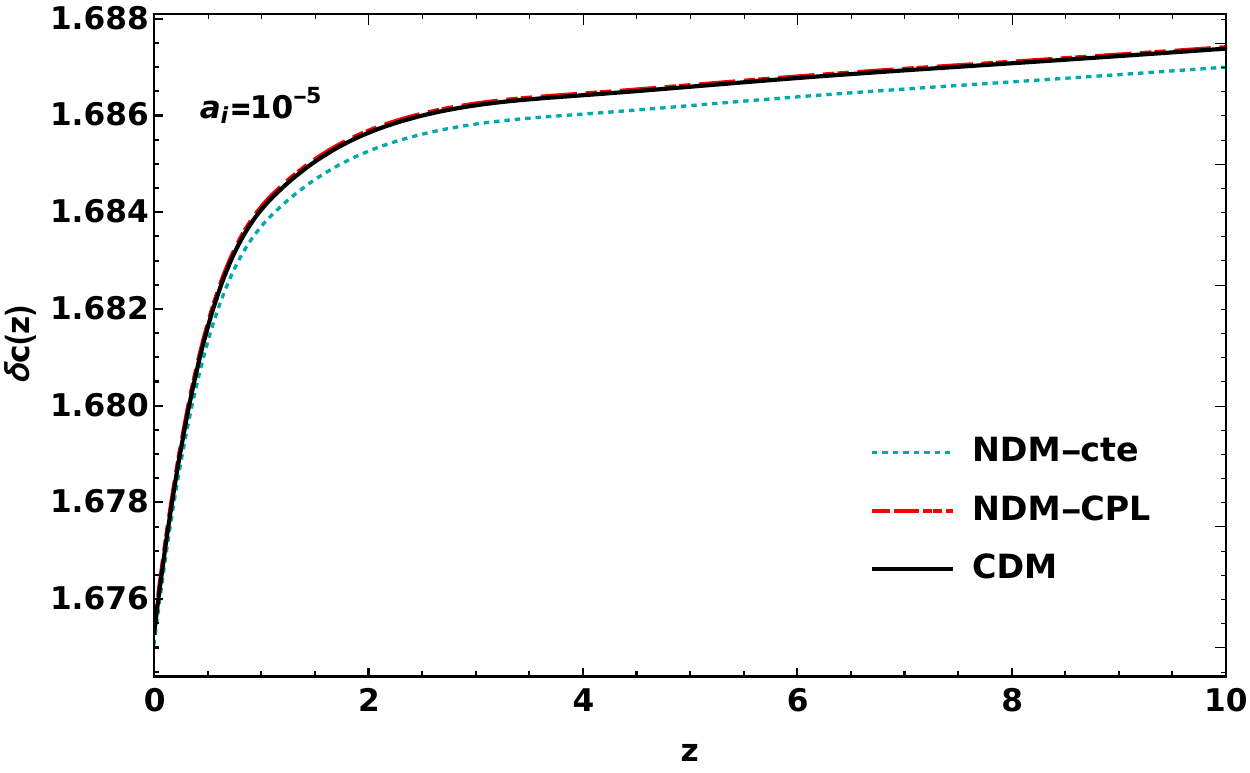}
	\includegraphics[height=4cm,width=8cm]{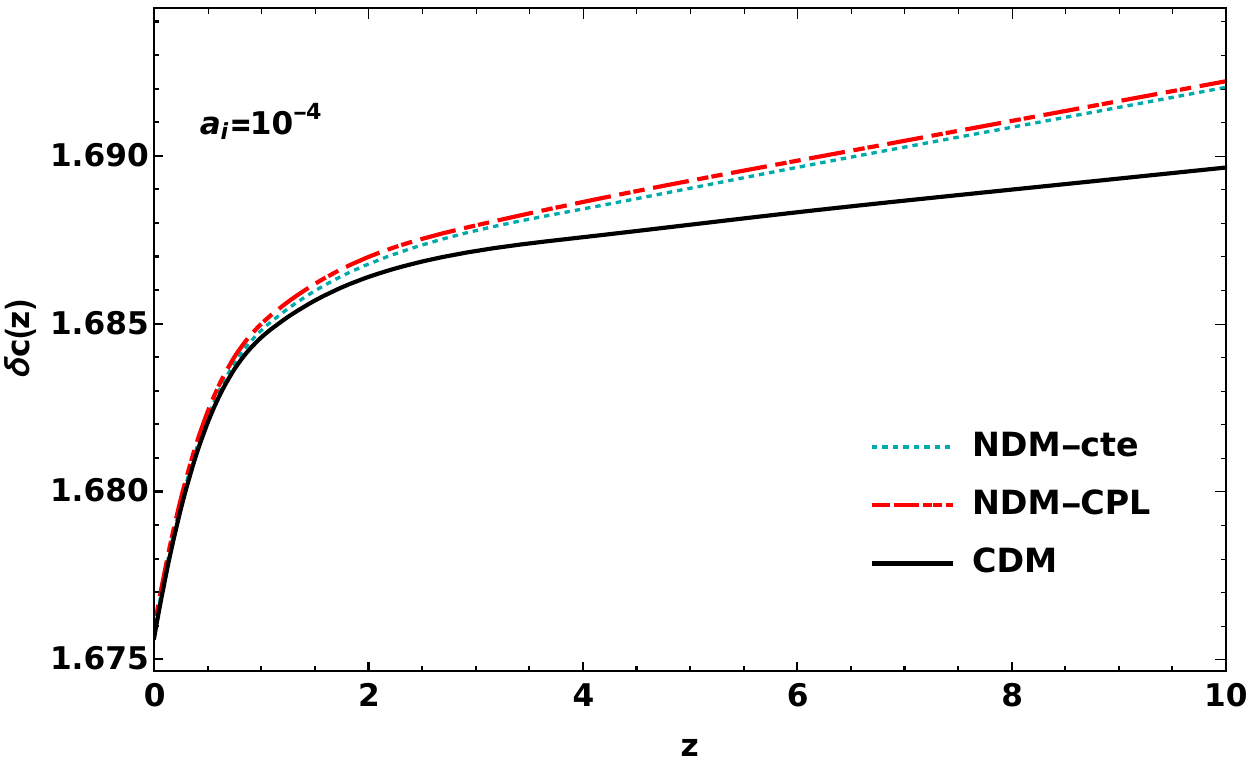}
	\caption{Critical collapse density $\delta_c$ as a function of redshift for different values of $a_i$.
	}
	\label{figdc}
\end{center}
\end{figure*}


\section{Nonlinear perturbations}
\label{sec:nonlin}

In this section, we study the nonlinear structure formation in the presence of a non-standard dark matter component. To investigate nonlinear perturbation theory and compute the formation rate of massive objects,  the simplest (semi-) analytical tool that can provide exact nonlinear results is the spherical collapse (SC) formalism. The SC method allows for explicit computation of nonlinear dynamics, providing a basis for evaluating several cosmological quantities such as halo mass functions and the probability distributions of the density contrast~\citep{Brax:2012sy}.

Now, we calculate two main quantities of the SC formalism, the critical density contrast at collapse, $\delta_c$, and the viral overdensity parameter, $\Delta_{{}_{\rm vir}}$, for different dark matter scenarios. $\delta_c$ is defined as the value of the linear density contrast at the redshift where the nonlinear density contrast diverges. The linear overdensity $\delta_c$ is used to calculate the mass function of dark matter halos in the Press-Schechter formalism, while the virial overdensity $\Delta_{\rm vir}$ determines the size of virialized halos.

To determine $\delta_c$ and $\Delta_{\rm vir}$, we follow the general approach outlined in~\citep{Pace:2014taa}. We seek the appropriate initial conditions,  $ \delta_{m,i} $, and its derivative $ \delta'_{m,i} $ to achieve highly nonlinear values of overdensities, $\delta\rightarrow\infty$, such as $ \sim 10^7 $, at a collapse redshift $ z_c $ using equation~\eqref{perm}. We rewrite the equation~\eqref{perm} in a linear regime as
\begin{eqnarray}\label{perl2}
&&\delta''_m+\left(\frac{3}{a}(1+c^2_s-w)+\frac{H'}{H}-\frac{w'}{1+w}\right)\delta'_m \nonumber \\
&&\qquad-\frac{3}{2a^2}\Omega_m(a)(1+w)(1+3c^2_s)\delta_m=0 \, .
\end{eqnarray}
Using the obtained initial conditions, we solve the linear equation~\ref{perl2} to calculate the linear overdensity $\delta_c=\delta_m(z=z_c)$. Since the initial scale factor for the numerical solution of the equations is not predetermined, we can consider two cases, $a_i=10^{-5}$ and $a_i=10^{-4}$. We find that for $a_i = 10^{-5}$, the critical density contrast $\delta_c$ as a function of redshift at collapse converges to the expected value of $1.686$ in the EdS Universe, as shown in Figure\ref{figdc}. Therefore, as mentioned in \citep{Pace:2017qxv}, accurate determination of initial condition will have significant consequences, so we focus on $a_i = 10^{-5}$ as the initial scale factor for the rest of the study. In both models, $\delta_c$ approaches a constant value at high redshifts, with $\delta{c,{\rm CDM}} < \delta{c,{\rm NDM-cte}} < \delta{c,{\rm NDM-CPL}}$. At high redshifts, the Universe is dominated by dark matter, whereas at lower redshifts, $\delta_c$ decreases and deviates from the EdS limit.

\begin{figure}
\begin{center}
	\includegraphics[height=4cm,width=8cm]{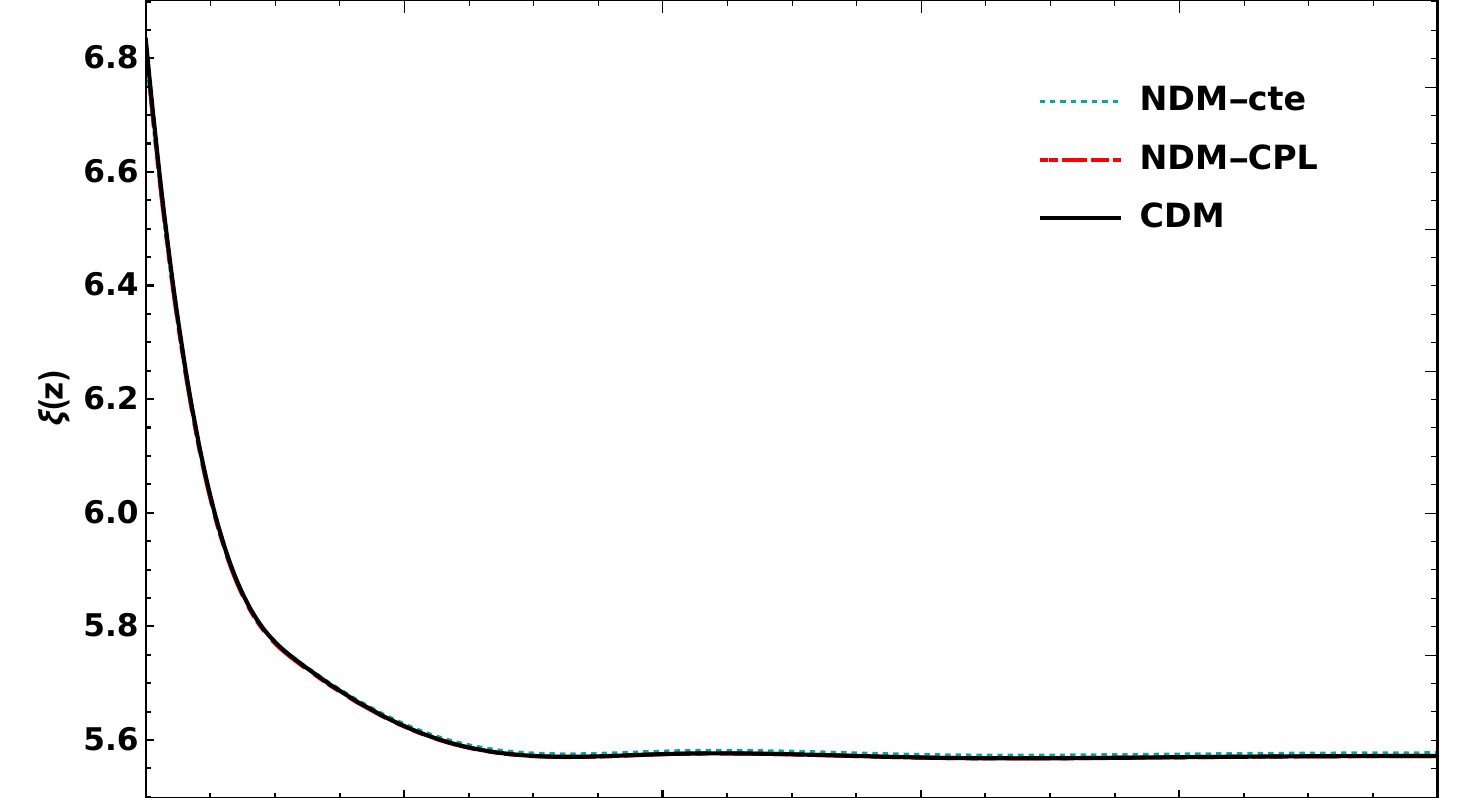}
	\includegraphics[width=8cm]{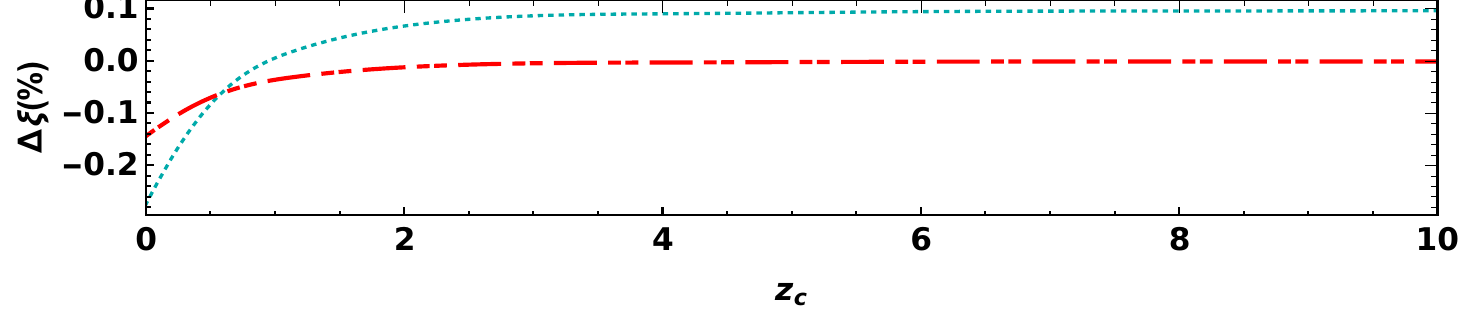}
	\caption{Evolution of $\xi$ as a function of collapse redshift, $z_c$, for NDM-cte and NDM-CPL models and their difference with CDM cosmology.}
	\label{fig-ksi}
\end{center}
\end{figure}

\begin{figure*}
\begin{center}
	\includegraphics[height=4cm,width=8cm]{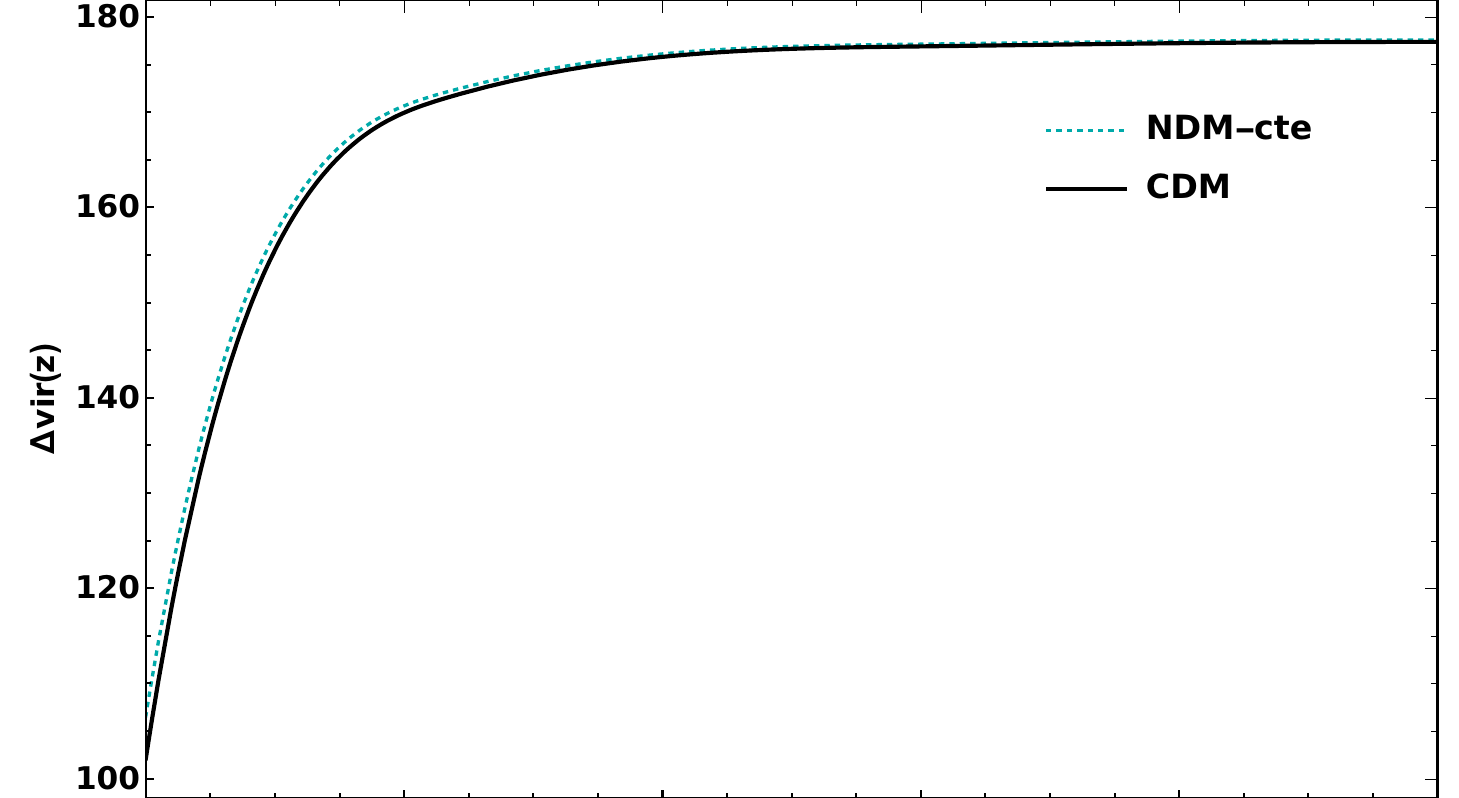}
	\includegraphics[height=4cm,width=8cm]{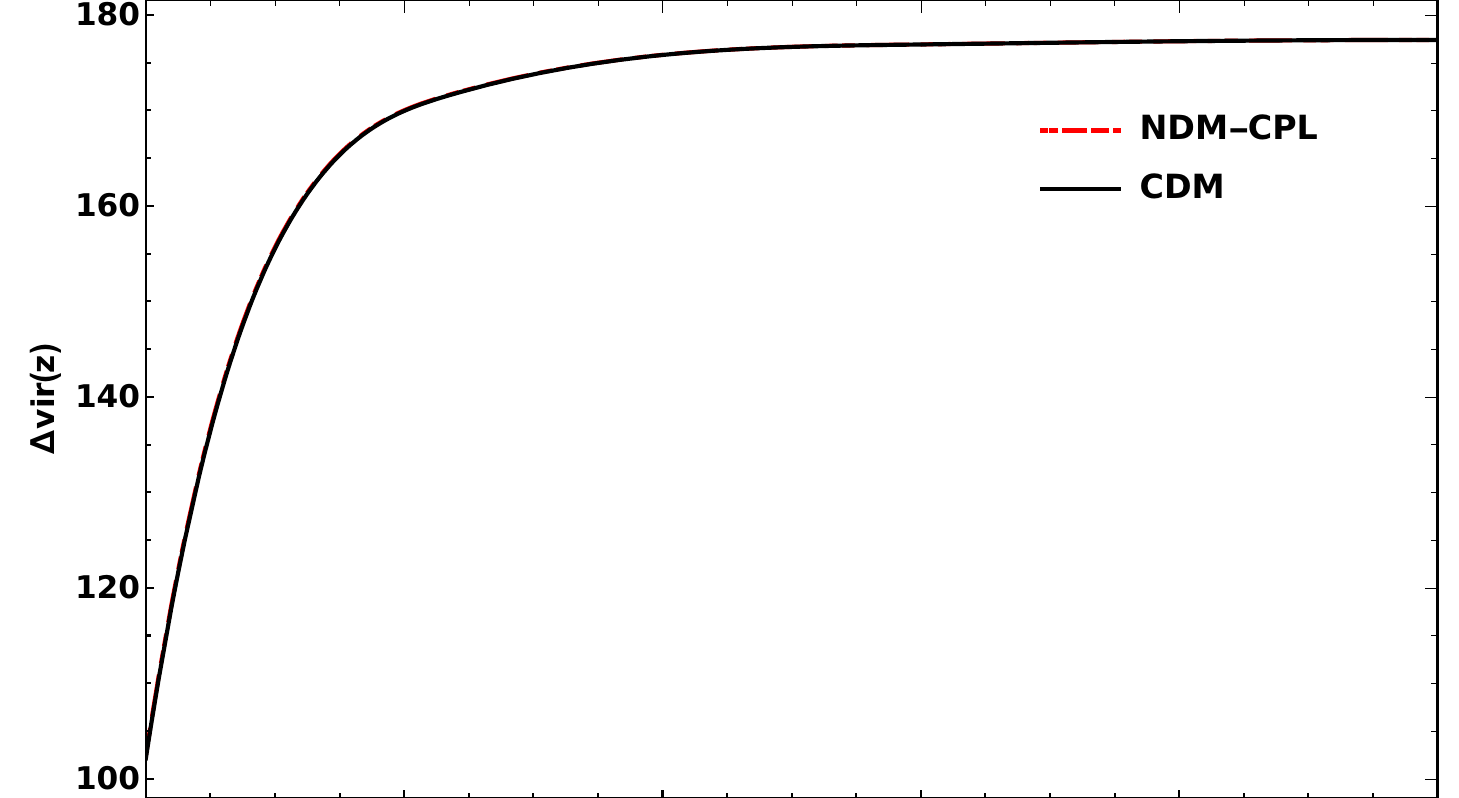}
	\includegraphics[width=8cm]{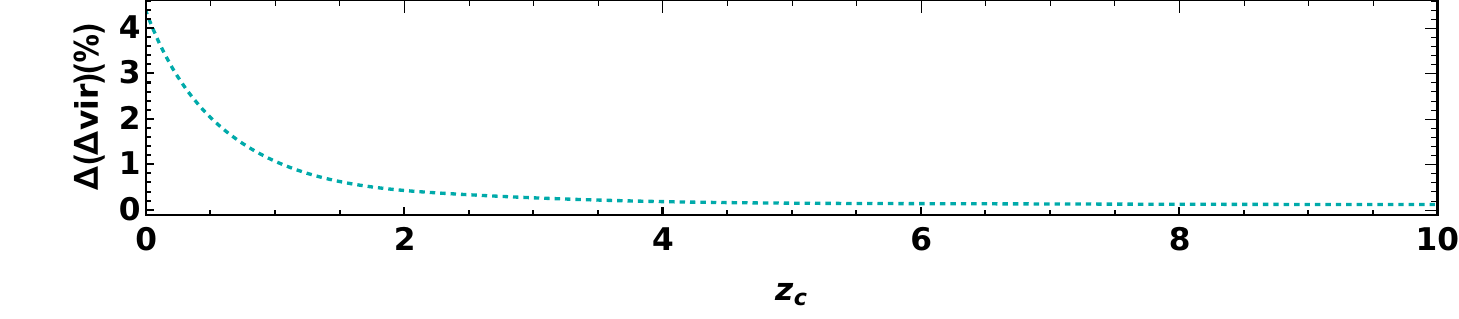}
	\includegraphics[width=8cm]{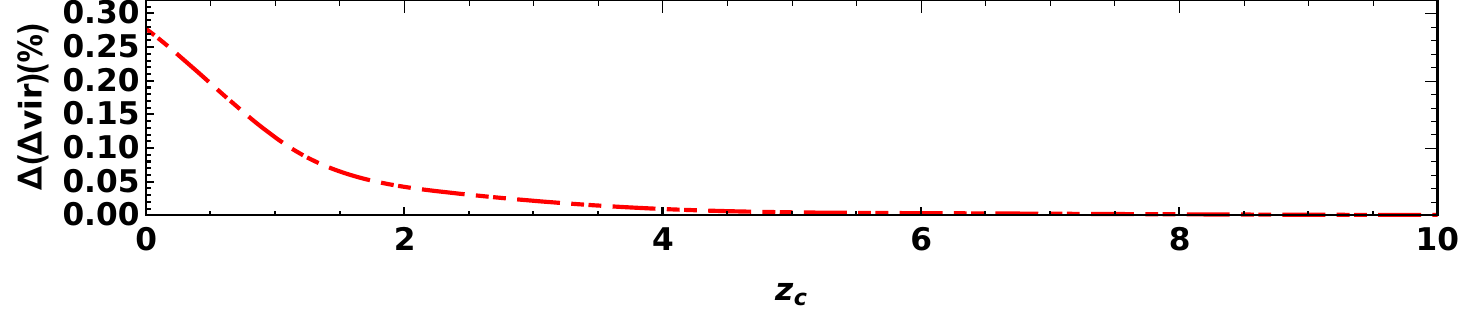}
	\caption{Evolution of the virial overdensity, $\Delta_{\rm vir}$, as a function of the collapse redshift, $z_c$, for the NDM-cte (left panels) and NDM-CPL (right panels) models and their difference with the CDM cosmology.}
	\label{fignon-cte}
\end{center}
\end{figure*}

Next, we investigate the virialization process of dark matter halos in the context of SC formalism. The overdensity of viralized halos is defined as $\Delta_{{}_{\rm vir}}=\xi(x/y)^3$, where $\xi$ is the amount of overdensity at the turnaround epoch, $ x $ is the scale factor normalized to turnaround scale factor ($ x=a/a_{{}_{\rm ta}} $ ), and $ y $ is the ratio between the virial radius and the turn-around radius ($y=R_{{}_{\rm vir}}/R_{{}_{\rm ta }}$)~\citep{Wang:1998gt}. In the preliminary EdS cosmology, we can easily obtain $ y=1/2 $, $\xi\approx 5.6$, and $\Delta_{{}_{\rm vir}}=178$ at any cosmic epoch of the Universe. Figure~\ref{fig-ksi} shows that in all the models, the turnaround overdensity $\xi$ converges to the fiducial value $\xi\approx 5.6$, which represents EdS cosmology at early times. Results for the virial overdensity $\Delta_{\rm vir}$ are presented in Figure~\ref{fignon-cte}. At low redshifts,  the virial overdensities in the dynamical DM and CDM models are lower than in the EdS Universe ($\Delta_{\rm vir} = 178$). This can be interpreted as the effect of DE on the process of virialization: DE prevents more collapse and, consequently, halos viralize at a larger radius with a lower density.

At the end of this section, we remember that although these differences are small, their combined effect enters exponentially in the evaluation of the halo mass function. Consequently, even a small difference can be amplified, leading to a noticeable difference, especially at the high-mass end, making it a valuable probe for cosmology~\citep{Pace:2019vrs}~\citep{Pace:2019vrs}.


\section{Halo mass function and cluster number count}
\label{sec:halomass}

As we know, the spherical collapse parameters cannot be observed directly, so it is more convenient to evaluate the comoving number density of virialized structures with masses in a certain range. This quantity is closely related to observations of structure formation. Therefore, in this section, we study the distribution of the number density of collapsed objects within a given mass range using a sophisticated formalism, named Press-Schechter, assuming alternatives to cold dark matter. Additionally, we consider the Sheth-Tormen (ST) formalism, a generalization of Press-Schechter, which offers improved accuracy in predicting halo mass functions. The mathematical formulations of the halo mass function are based on the assumption of a Gaussian distribution of the matter density field. The comoving number densities of virialized halos with masses in the range of $M$ and $M+dM$ in terms of collapse redshift $z$ is given by:
\begin{equation}
\frac{dn(z,M)}{dM}=-\frac{\rho_{m,0}}{M}\frac{d\ln\sigma(z,M)}{dM}f(\sigma(z,M)) \, ,
\end{equation}
where $\rho_{m,0}$ is the present-day matter energy density and $ f(\sigma) $ is  the mass function\citep{Bond1991}. A traditional Gaussian distribution function, $ f(\sigma) $, was proposed by Press and Schechter (PS)~\citep{Press:1973iz} as
\begin{equation}\label{fs}
f(\sigma)=\sqrt{\frac{2}{\pi}}\frac{\delta_c(z)}{\sigma(z,M)}\exp\left[-\frac{\delta_c^2(z)}{2\sigma^2(z,M)}\right] \, ,
\end{equation}
here $\sigma$, the root mean square (rms) of the mass fluctuations in spheres of radius $R$ containing the mass $M$~\citep{Press:1973iz}, obtained from
\begin{equation}
	\sigma^2(z,R) = \frac{1}{2\pi^2}\int_{0}^{\infty} P(k,z) W^2(kR)k^2dk.
\end{equation}
where  $W(kR)$, the Fourier transform of a spherical top-hat profile with radius R is defined by
$W(kR) = 3[\sin(kR)- kR \cos(kR)]/(kR)^3$ and radius is given by $R = [3M_h/(4\pi \rho_m)]1/3$ with $M_h$ being the
mass of the halo and $\rho_m$ the mean matter density of the universe at the present time.
 $P(k,z)$ is the linear power spectrum of density fluctuations. We use the procedure presented in  ~\citep{Viana:1995yv} to calculate $\sigma(z, M)$ in the vicinity of $8h^{-1}$ Mpc.
The function~\eqref{fs} works well for predicting the number density of virialized halos with moderate mass but fails at estimating high abundance for low-mass structures and low abundance for high-mass objects. Hence, in this work, we use the Sheth-Torman (ST) mass function, which is in better agreement with simulations across both low and high mass tails, given by~\citep{Sheth:1999mn}
\begin{align}
	&f(\sigma)=A\sqrt{\frac{2a}{\pi}}\left[1+\left(\frac{\sigma^2(z,M) }{a\delta_c^2(z)}\right)^{p}\right]\left(\frac{ \delta_c(z)}{\sigma(z,M)}\right)\times\nonumber\\
	&\qquad\qquad\qquad\exp\left[-\frac{a\delta_c^2(z)}{2\sigma^2(z,M)}\right]. \,
\end{align}
Here, $A=0.3222$, $a=0.707$, and $p=0.3$. Note that the PS mass function is recovered in the limit $A = a\rightarrow1$ and $p\rightarrow0$. As we see, the quantities $\delta_c$ and mass variance $\sigma$ depend strongly on the cosmological model considered.\\ 
According to the above formulation, the number density of virialized halos above a specified mass $M$ at collapse redshift $z$ is given by \cite{Abramo:2008ip,Malekjani:2016mtm}
\begin{equation}
n(>M_{{}_{\rm halo}},z)=\int_{M_{{}_{\rm halo}}}^{\infty}\frac{dn(z,M)}{dM}dM \, .
\end{equation}
Practically, we fix the upper limit of integration as $M = 10^{18} M_{\odot} h^{-1}$, as such a gigantic structure could not be observed. This quantity is of interest in high redshift galaxy observations.

\begin{figure*}
\begin{center}
	\includegraphics[height=4cm,width=8cm]{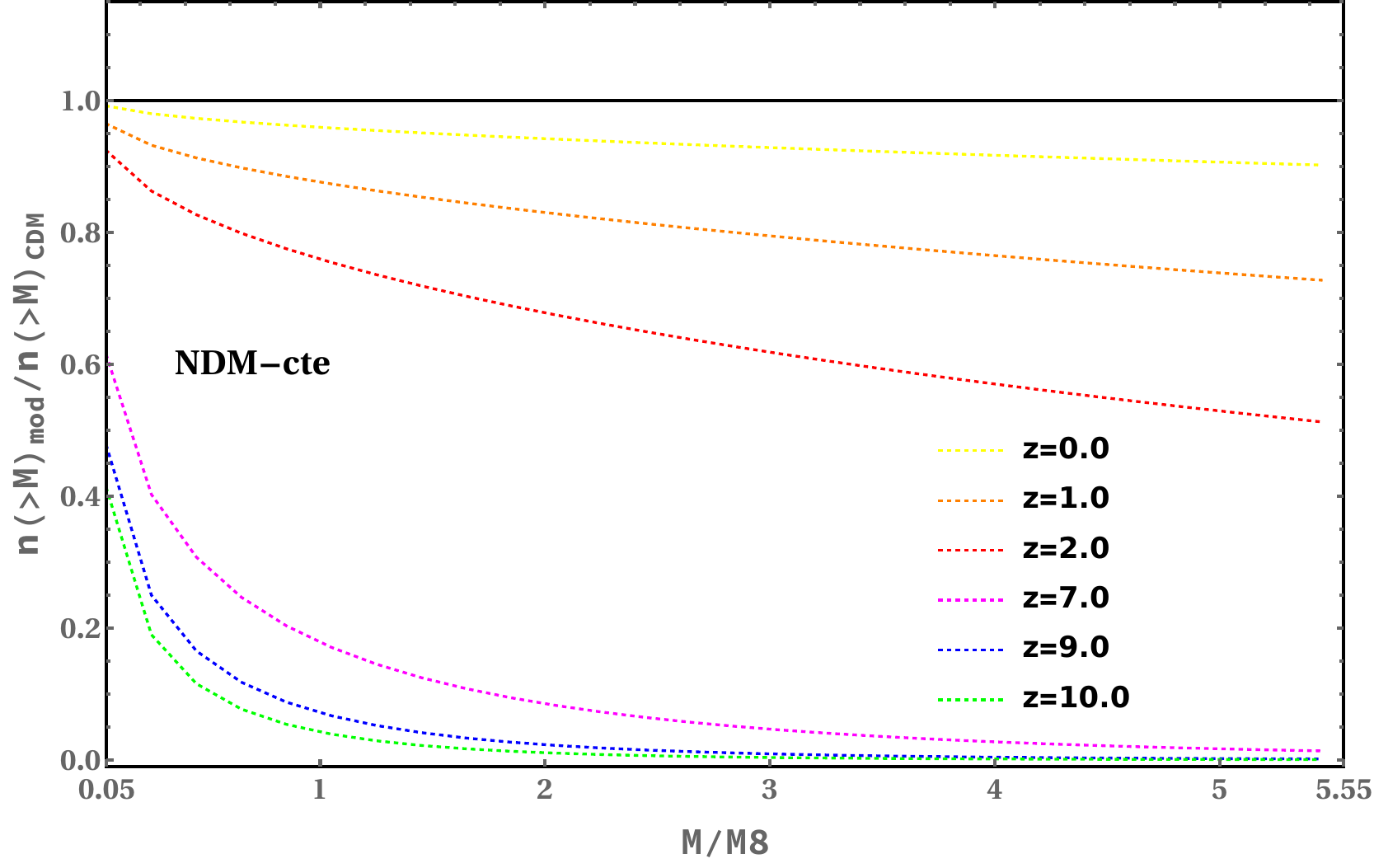}
	\includegraphics[height=4cm,width=8cm]{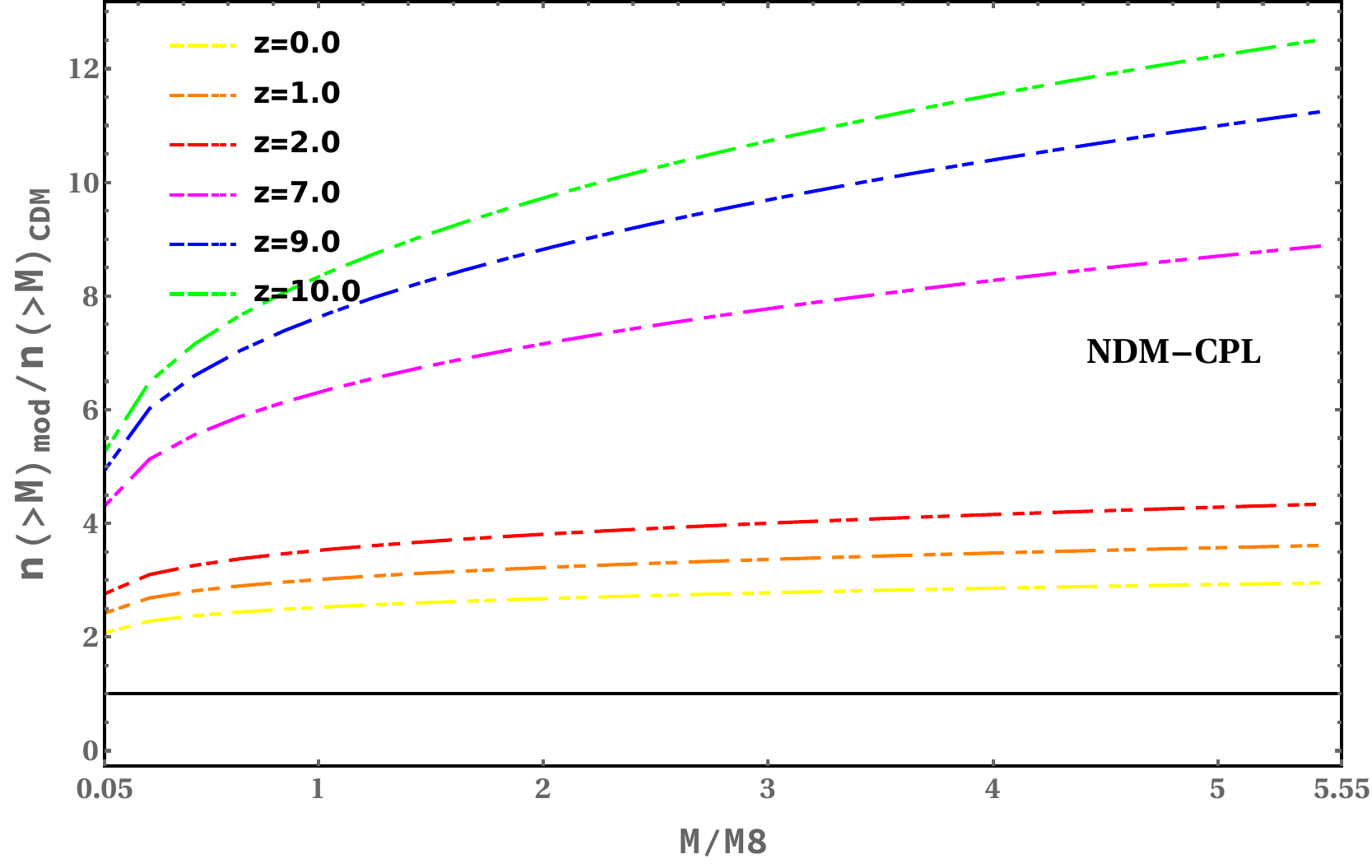}

	\caption{Ratio of the number of virialized halos above a certain mass $(>M)$ at $z=0$  until $z=10$ for the different dark matter models to the CDM cosmology. Line styles and colors are shown in the legends.}
	\label{fig5}
\end{center}
\end{figure*}

In Figure~\ref{fig5}, we show the ratio of the predicted number of halos above a certain mass $M$ normalized to the CDM model for two sets of cosmic redshifts: $z = 0, 1,2$ (low redshifts) and $z=7,9,10$ (high redshifts) in the mass range $10^{13} M_{\odot}\ll M\ll 10^{15}M_{\odot}$. As we observe in all cases, at higher redshifts, differences in halo numbers between alternative dark matter models and the CDM Universe appear even at low masses. In all redshifts, the NDM-CPL model predicts a higher number of virialized objects compared to the NDM-cte model. Thus, the additional characteristics of the dark matter sector significantly impact the observables, affecting both the linear and nonlinear evolution of structure formation through the halo mass function.

The cumulative halo mass density, $\rho_m(>M_{\rm halo},z)$, is another closely related quantity given by
\begin{equation}
\rho_m(>M_{{}_{\rm halo}},z)=\int_{M_{{}_{\rm halo}}}^{\infty}M\frac{dn(z,M)}{dM}dM \, .
\end{equation}
Assuming the largest stellar content a halo can have is its cosmic allotment of baryons, this converts directly to upper limits on the statistics of galaxies as $M_{\star, \max} = \epsilon f_b M_{\rm halo}$, where $f_b \equiv \frac{\Omega_b}{\Omega_m}$ is the cosmic baryon fraction and $\epsilon \leq 1$ is the efficiency of converting gas into stars.

A particularly useful quantity for JWST observations is the cumulative comoving mass density of stars contained in galaxies more massive than $M_{\star}$, $\rho_{\star}(>M_{\star},z)$, defined as \citep{Boylan-Kolchin:2022kae}:
\begin{equation}
\rho_\star(>M_\star,z)=\epsilon f_b\int_{M_\star/(\epsilon f_b)}^{\infty}M\frac{dn(z,M)}{dM}dM \, .
\end{equation}
We use this quantity to evaluate the viability of the CDM model. In the following section, we will explore how JWST observations can potentially discriminate between different dark matter models.
\begin{figure*}
\begin{center}
	\includegraphics[height=4cm,width=5cm]{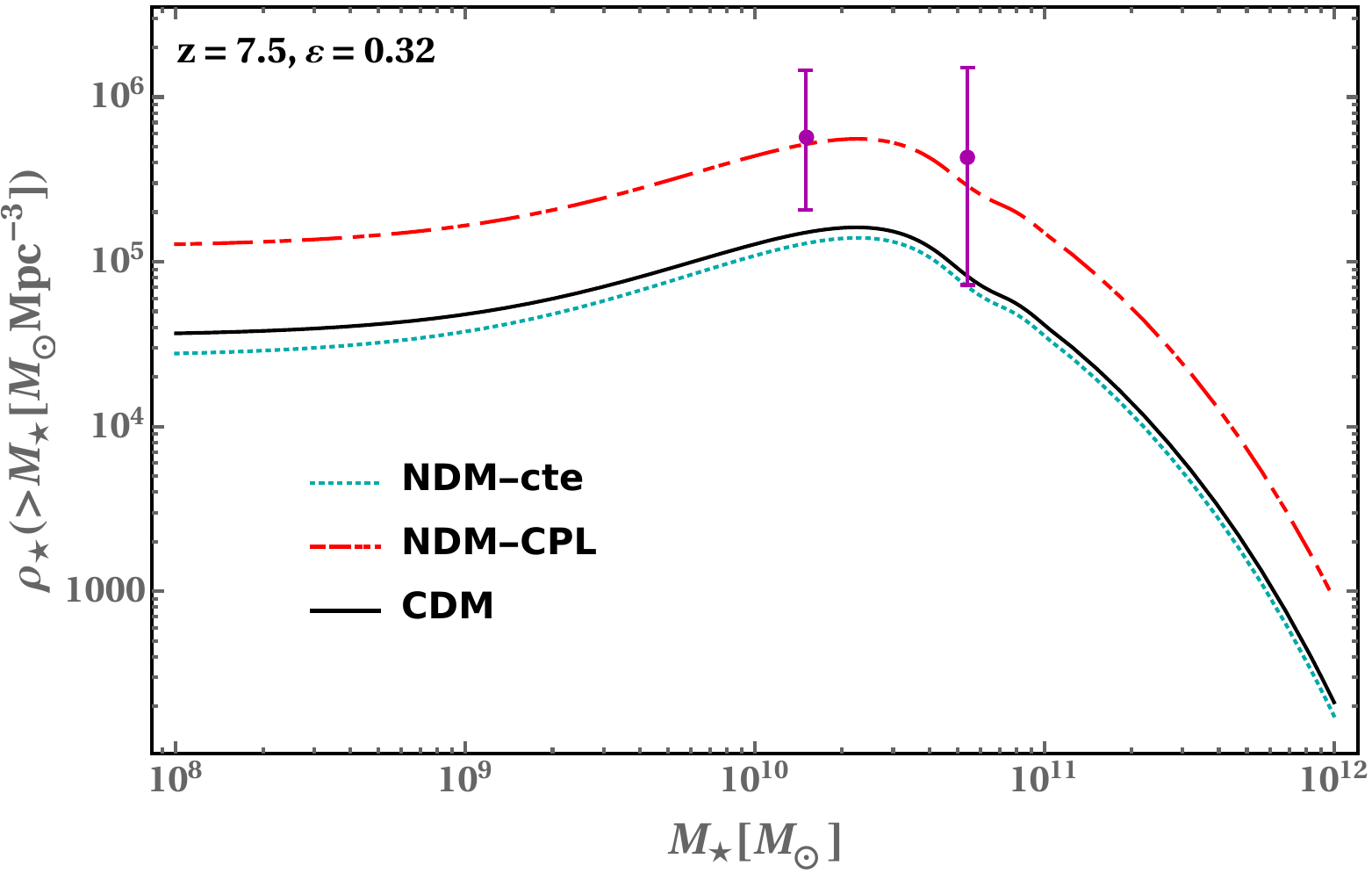}
	\includegraphics[height=4cm,width=5cm]{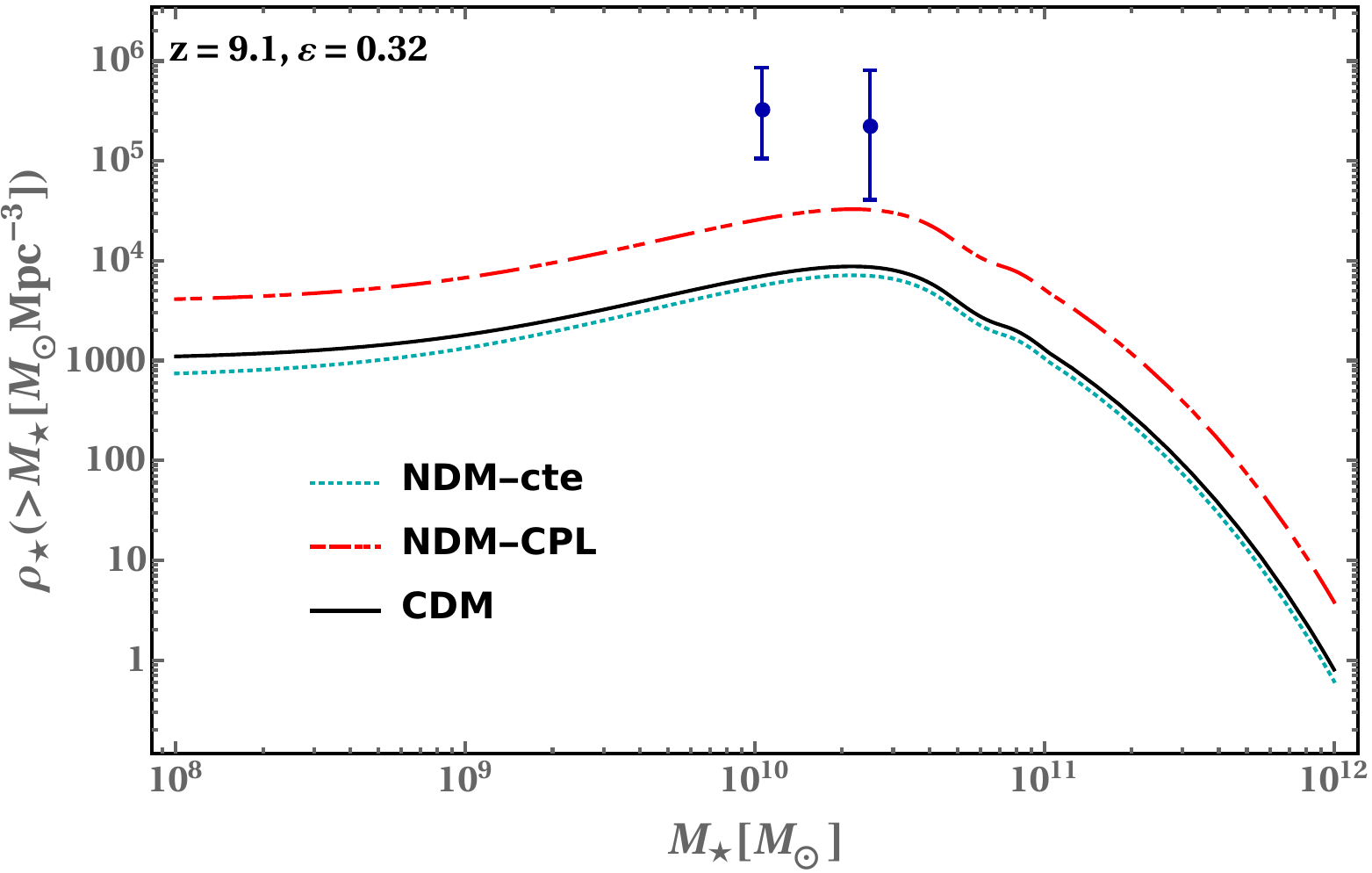}
	\includegraphics[height=4cm,width=5cm]{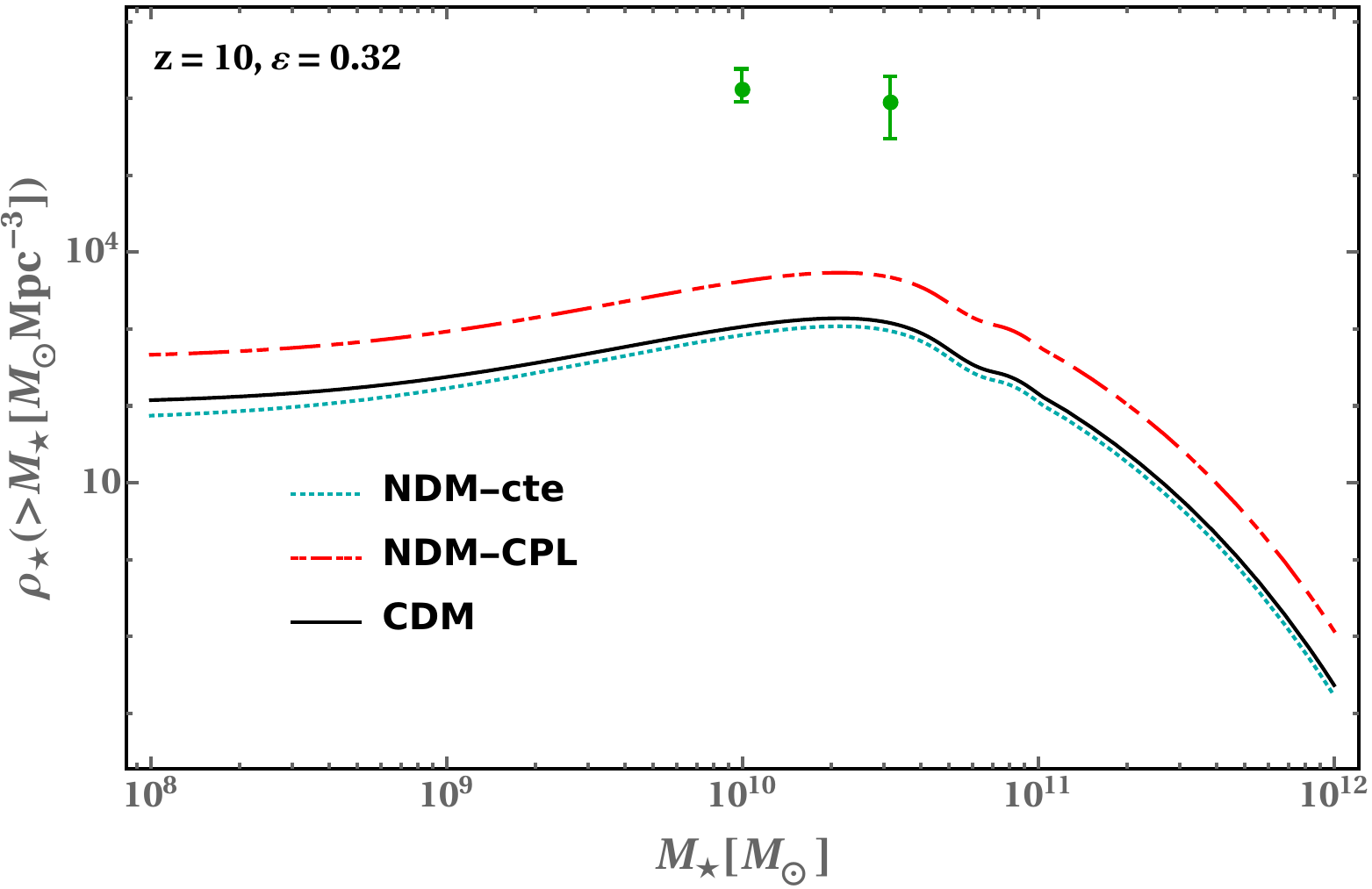}
	\includegraphics[height=4cm,width=5cm]{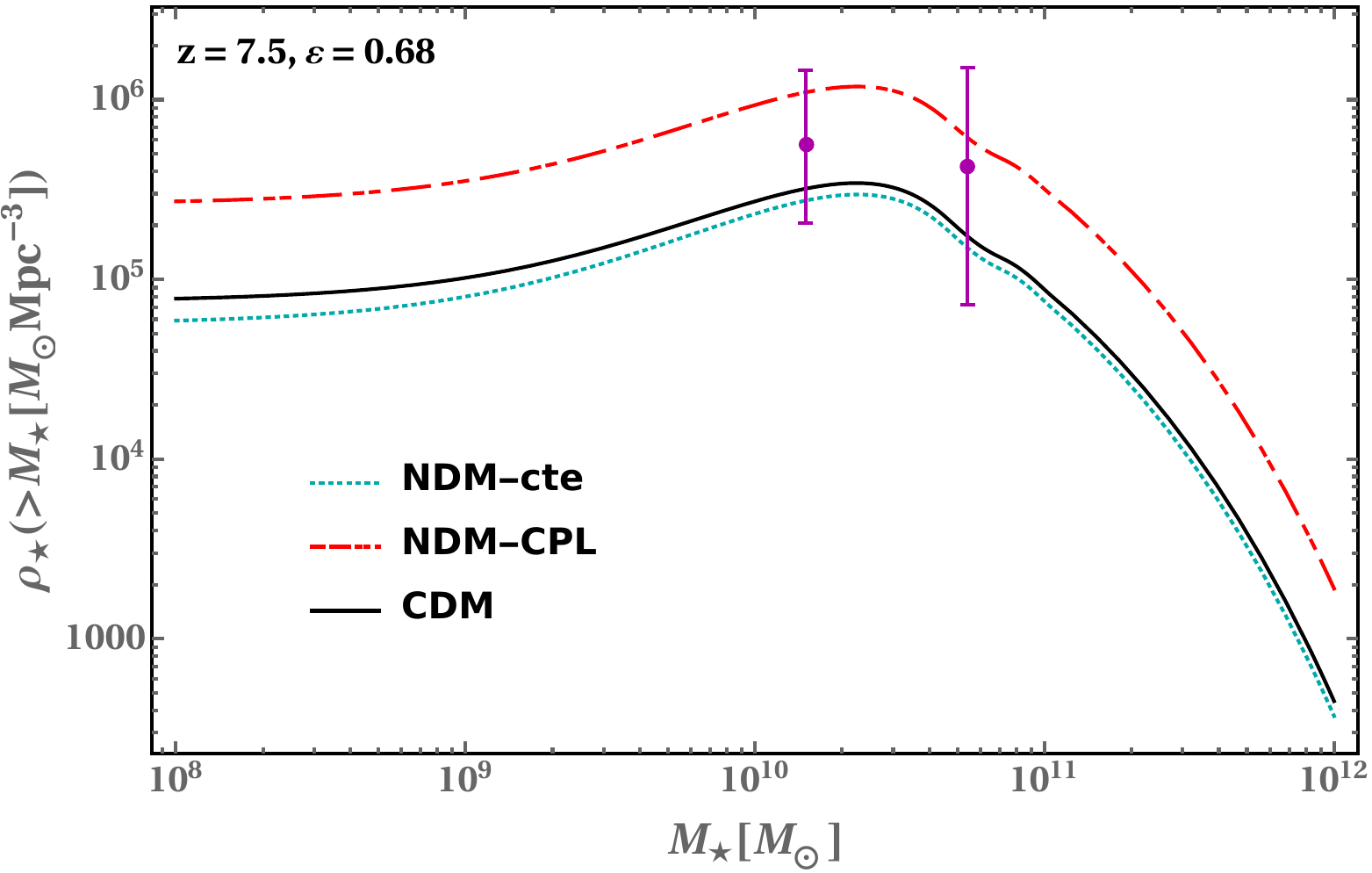}
	\includegraphics[height=4cm,width=5cm]{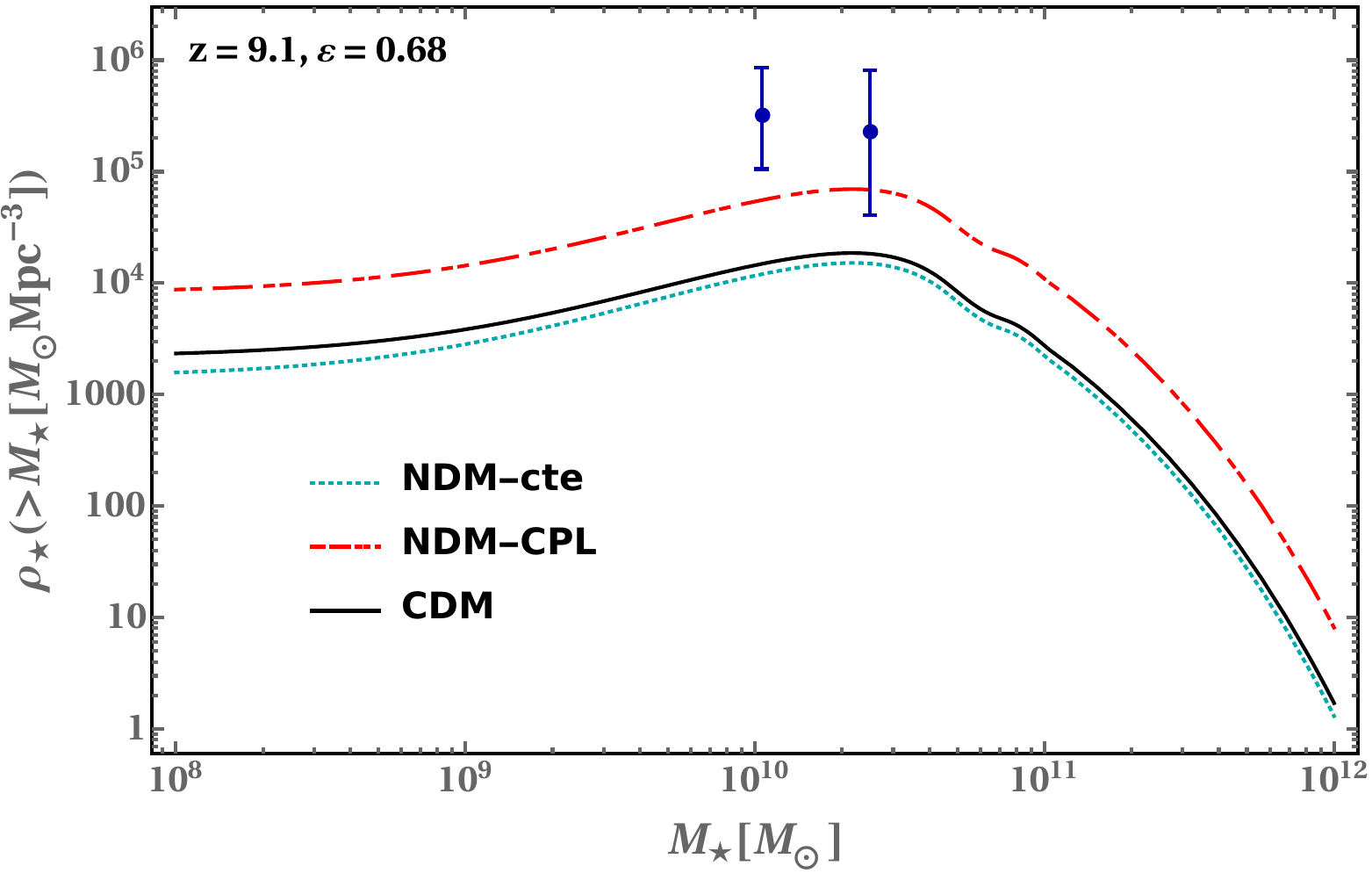}
	\includegraphics[height=4cm,width=5cm]{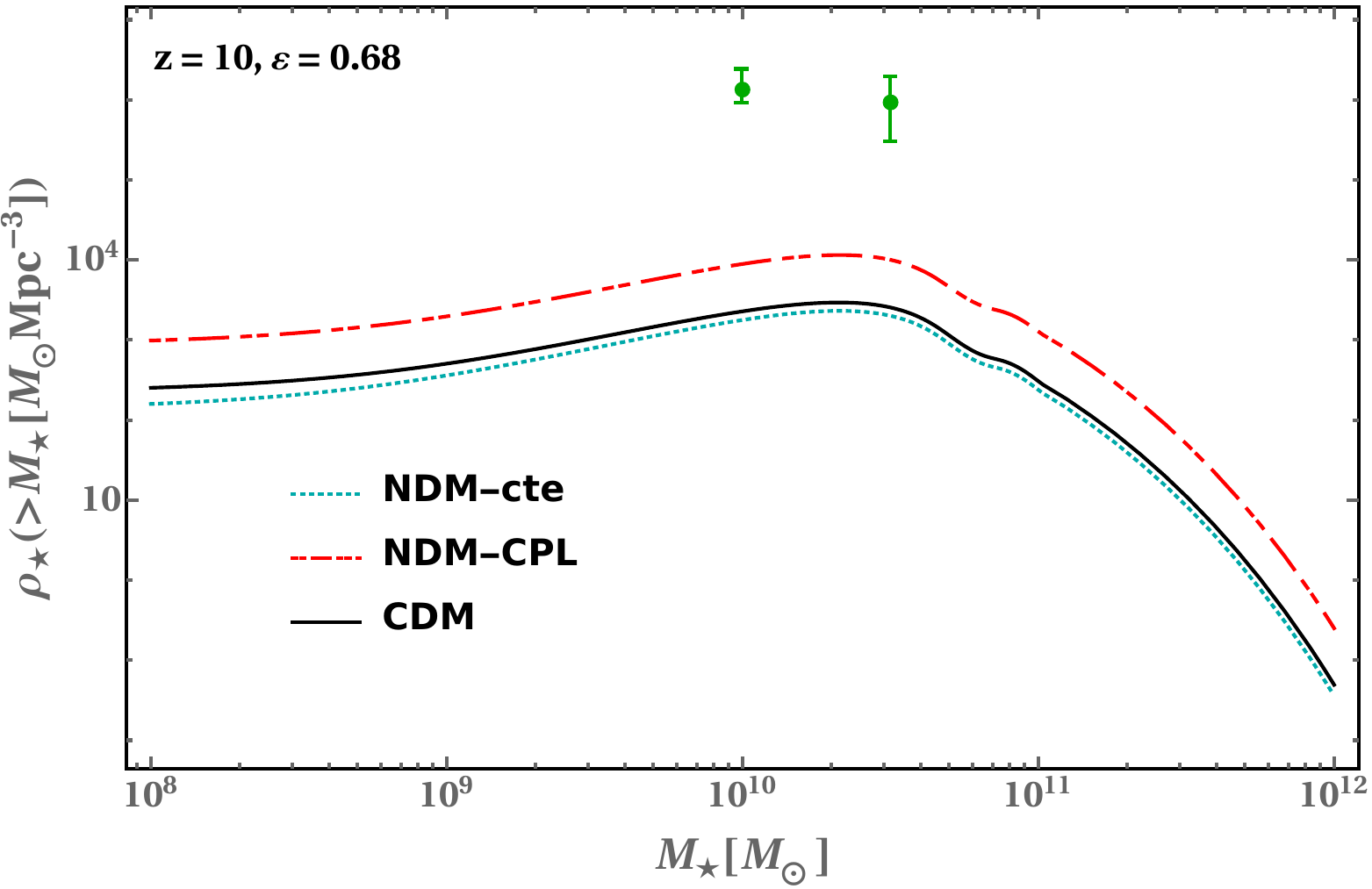}
	\includegraphics[height=4cm,width=5cm]{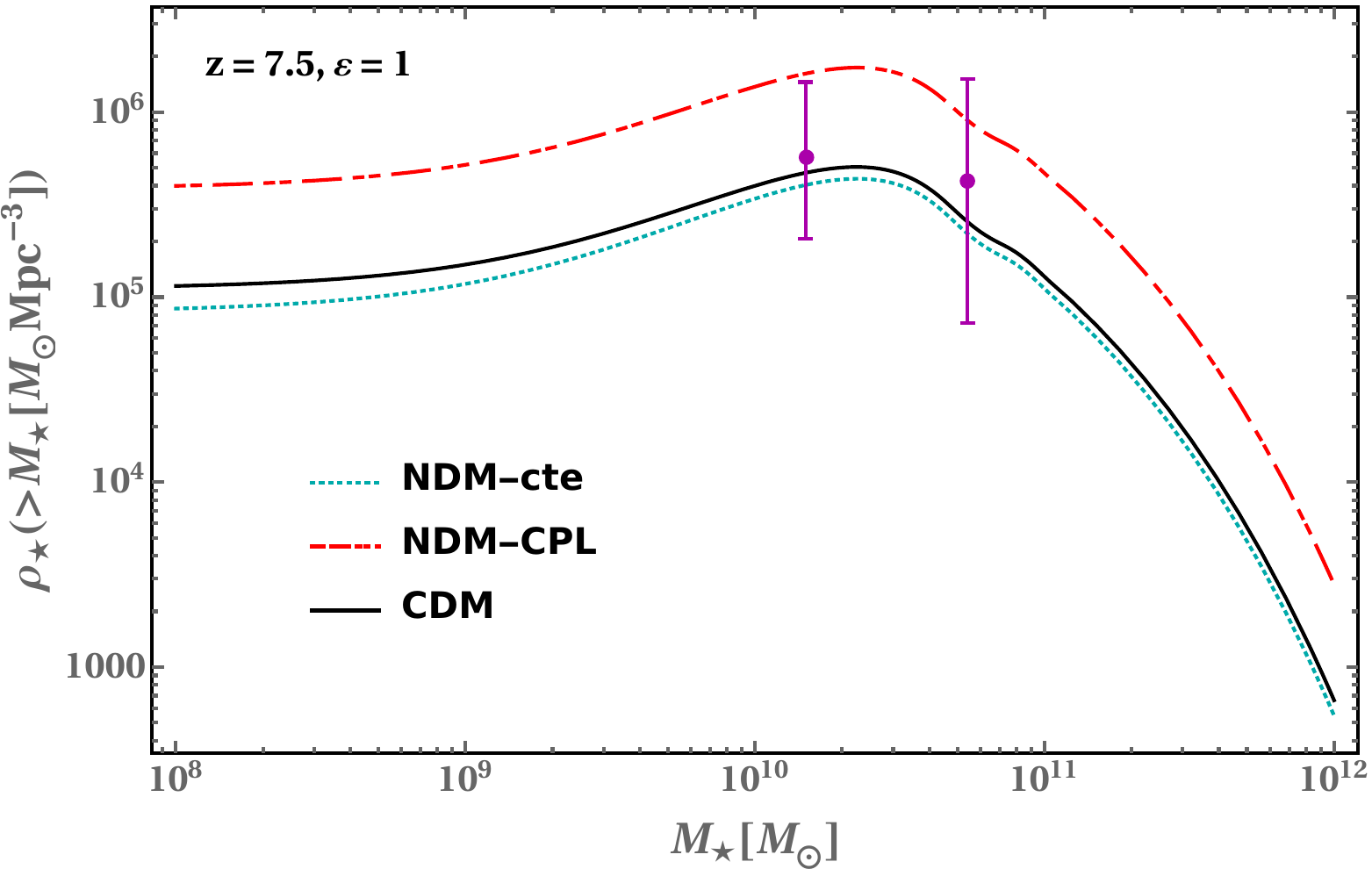}
	\includegraphics[height=4cm,width=5cm]{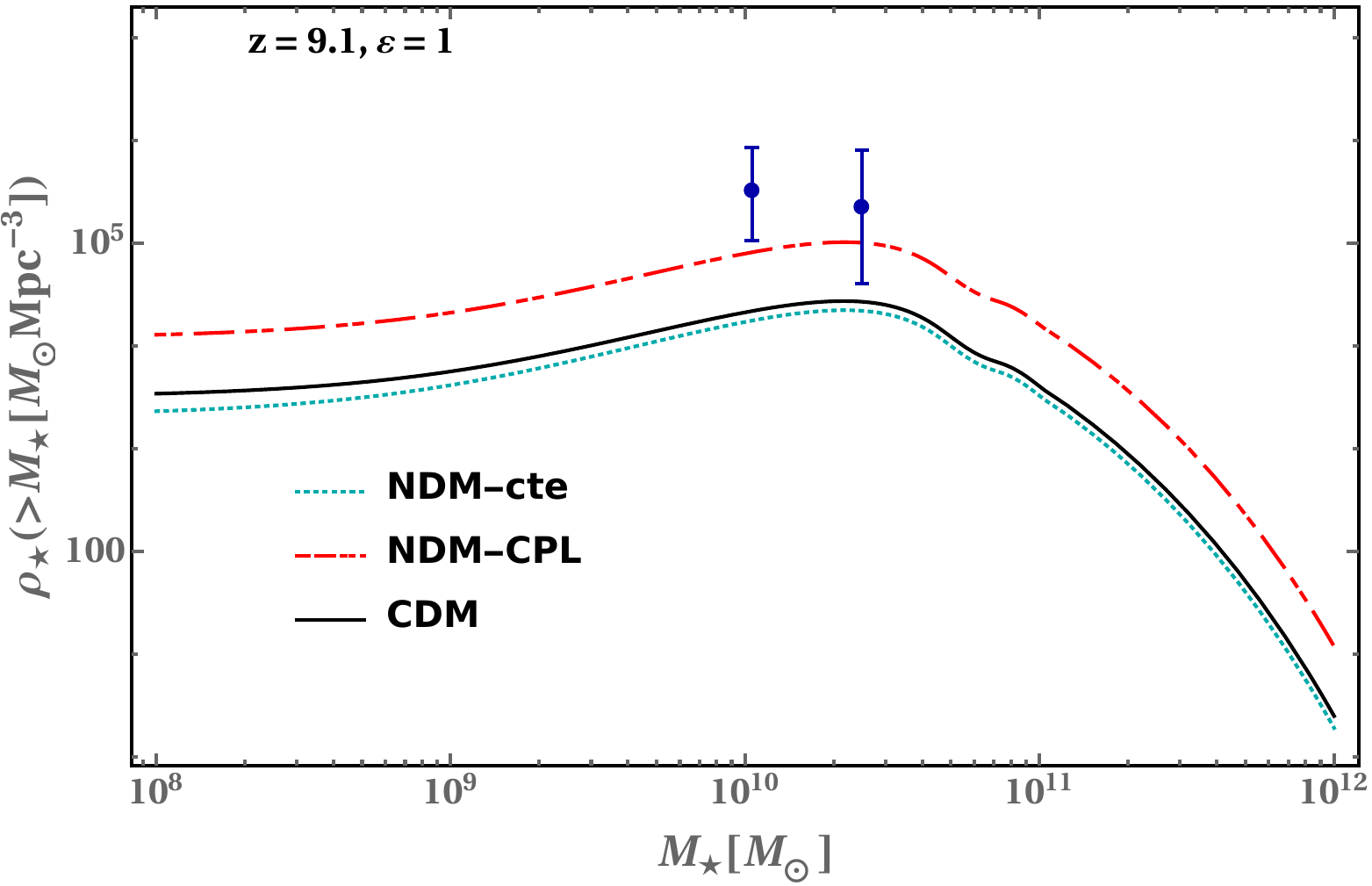}
	\includegraphics[height=4cm,width=5cm]{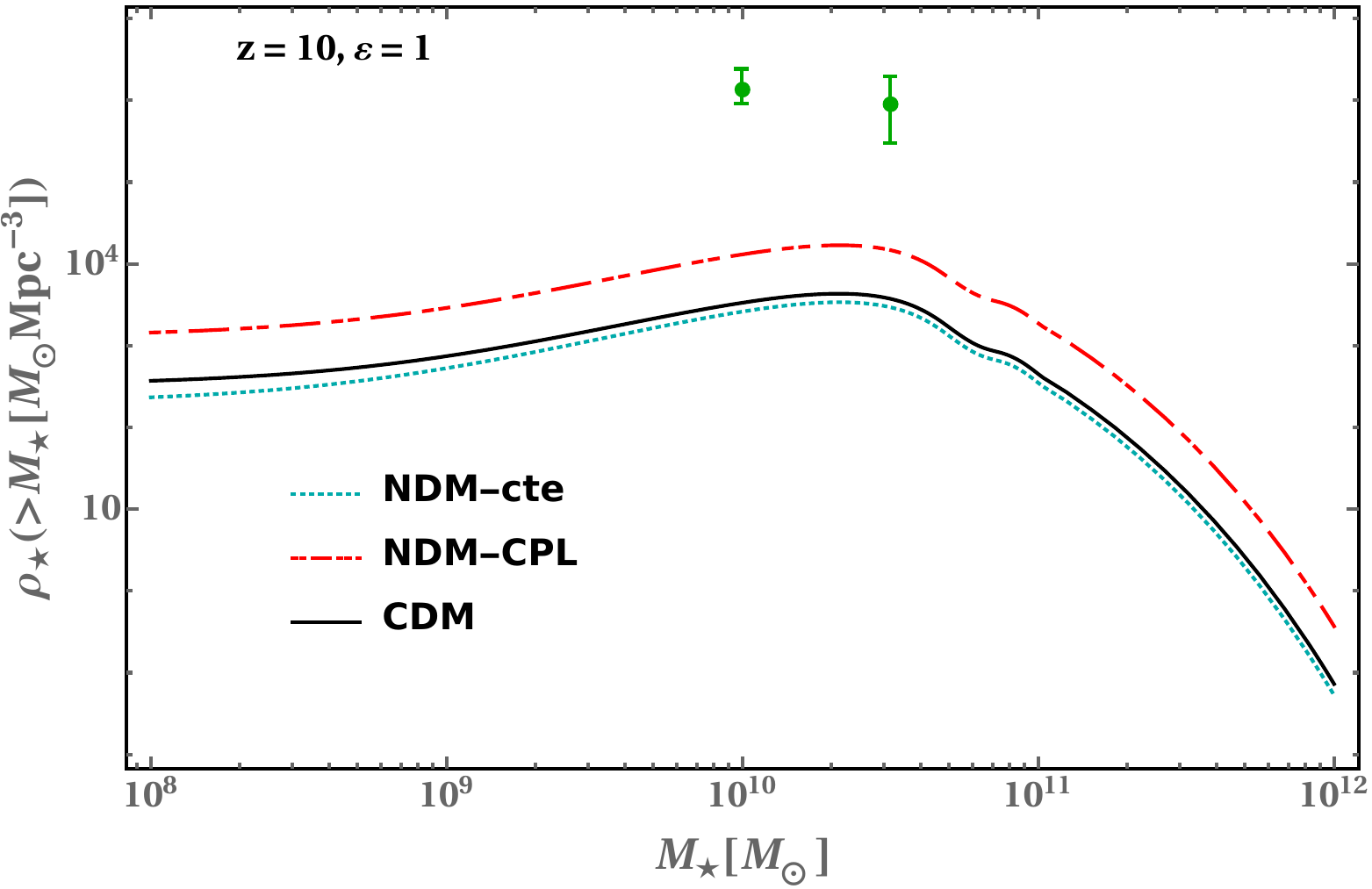}
	\caption{The comoving stellar mass density contained within galaxies more massive than $M_\star$ at $z = 7.5$ (left panel), $z = 9.1$ (middle panel), and $z = 10$ (right panel) for various values of the assumed conversion efficiency $\epsilon = 0.32, 0.68$ and $1$ of a halo's cosmic allotment of baryons into stars. The cosmological parameters are fixed according to their best values given in the Table~\ref{best}.}
	\label{fig6}
\end{center}
\end{figure*} 


\section{Comparison with the JWST observations}
\label{sec7}

Galaxies with stellar masses up to $\sim10^{11}M_\odot$ have been detected approximately one billion years after the Big Bang, up to redshift $z \sim 6$. Detecting massive galaxies at earlier times has been challenging  (known as the Balmer break region) since accurate mass estimates require observations at wavelengths beyond 2.5 $\mu$m.  On the other hand, studying high-redshift galaxies is crucial for understanding the formation and evolution of early galaxies and the history of cosmic reionization.
\begin{figure*}
\begin{center}
	\includegraphics[height=4cm,width=5cm]{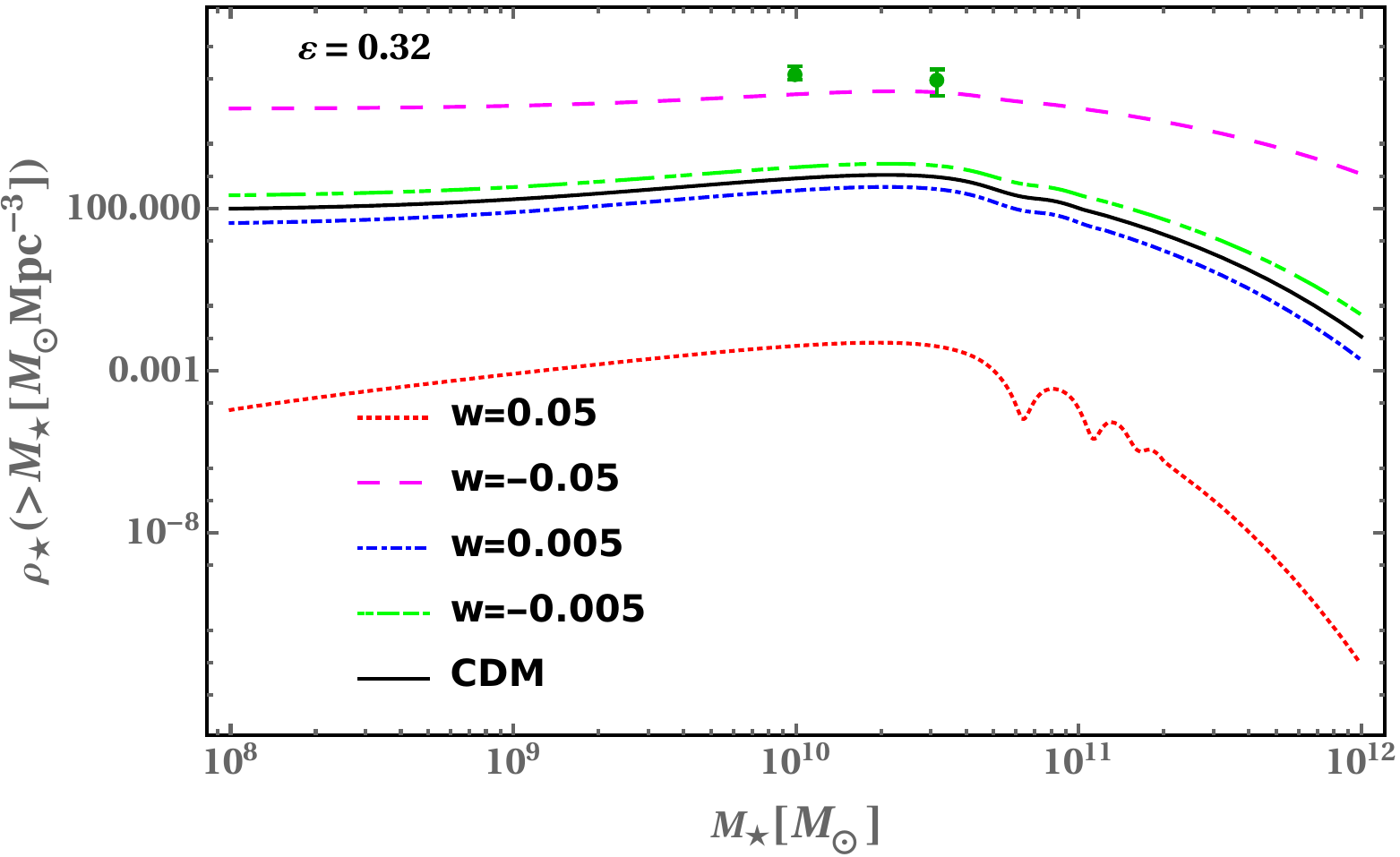}
	\includegraphics[height=4cm,width=5cm]{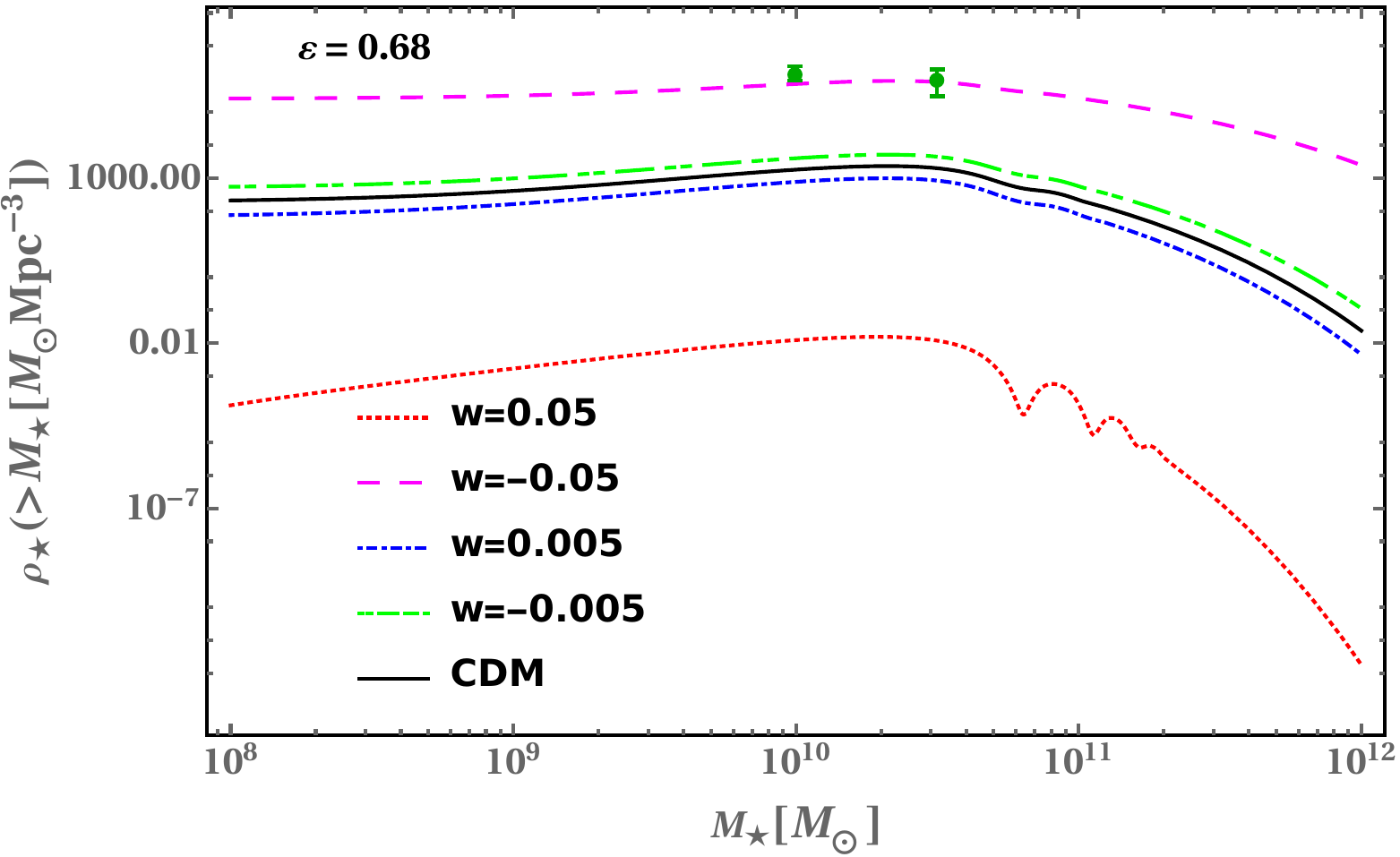}
	\includegraphics[height=4cm,width=5cm]{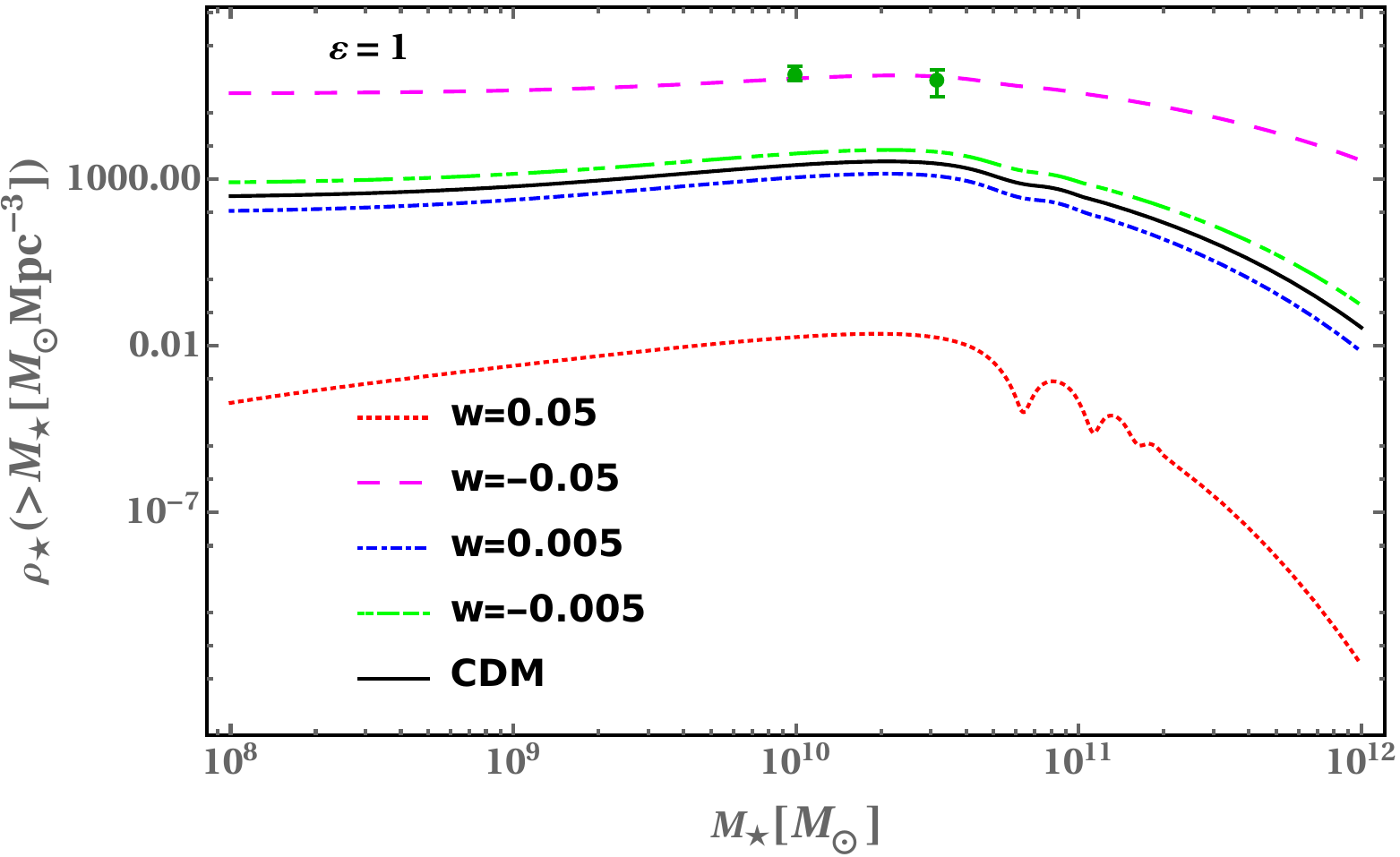}
	\includegraphics[height=4cm,width=5cm]{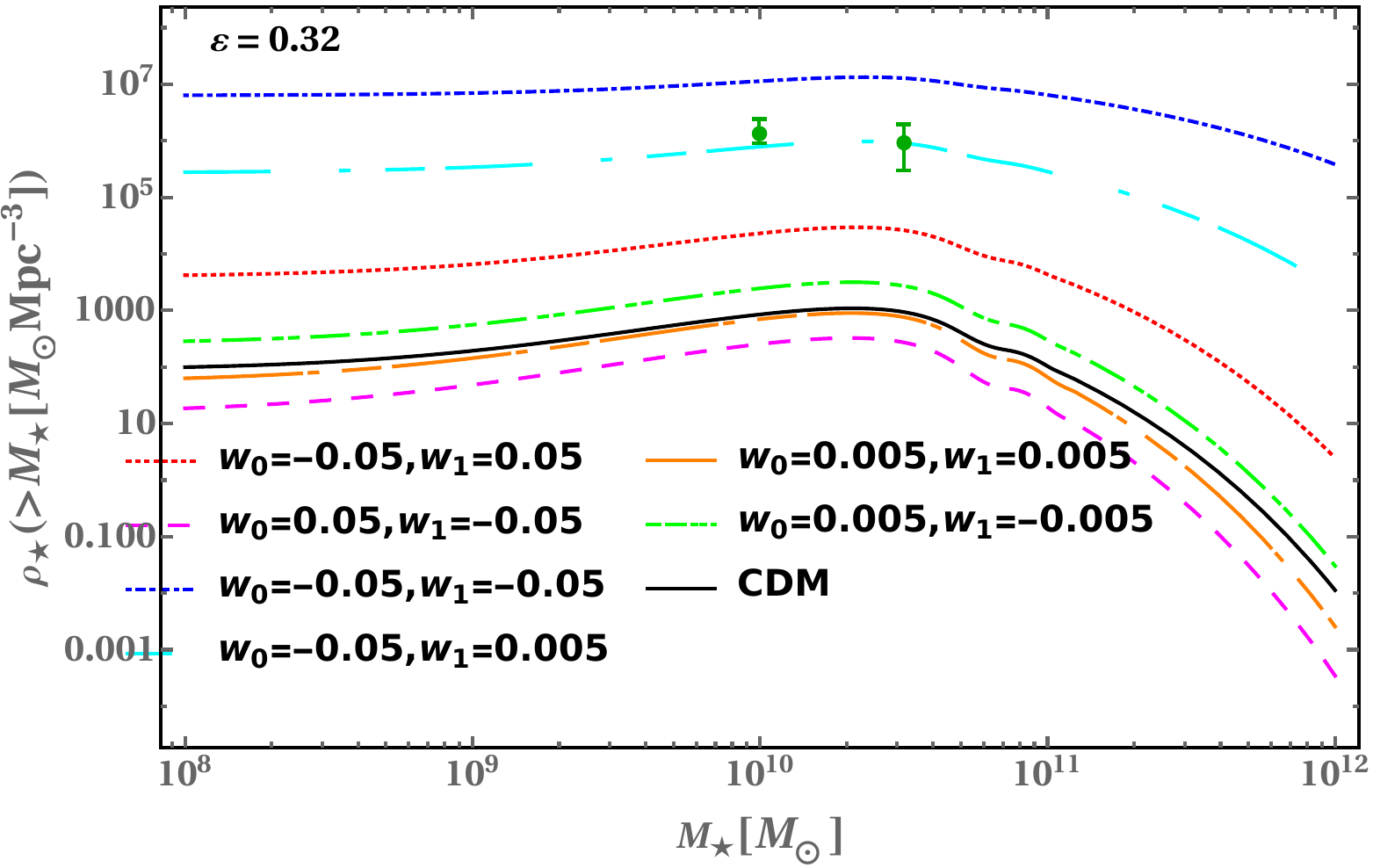}
	\includegraphics[height=4cm,width=5cm]{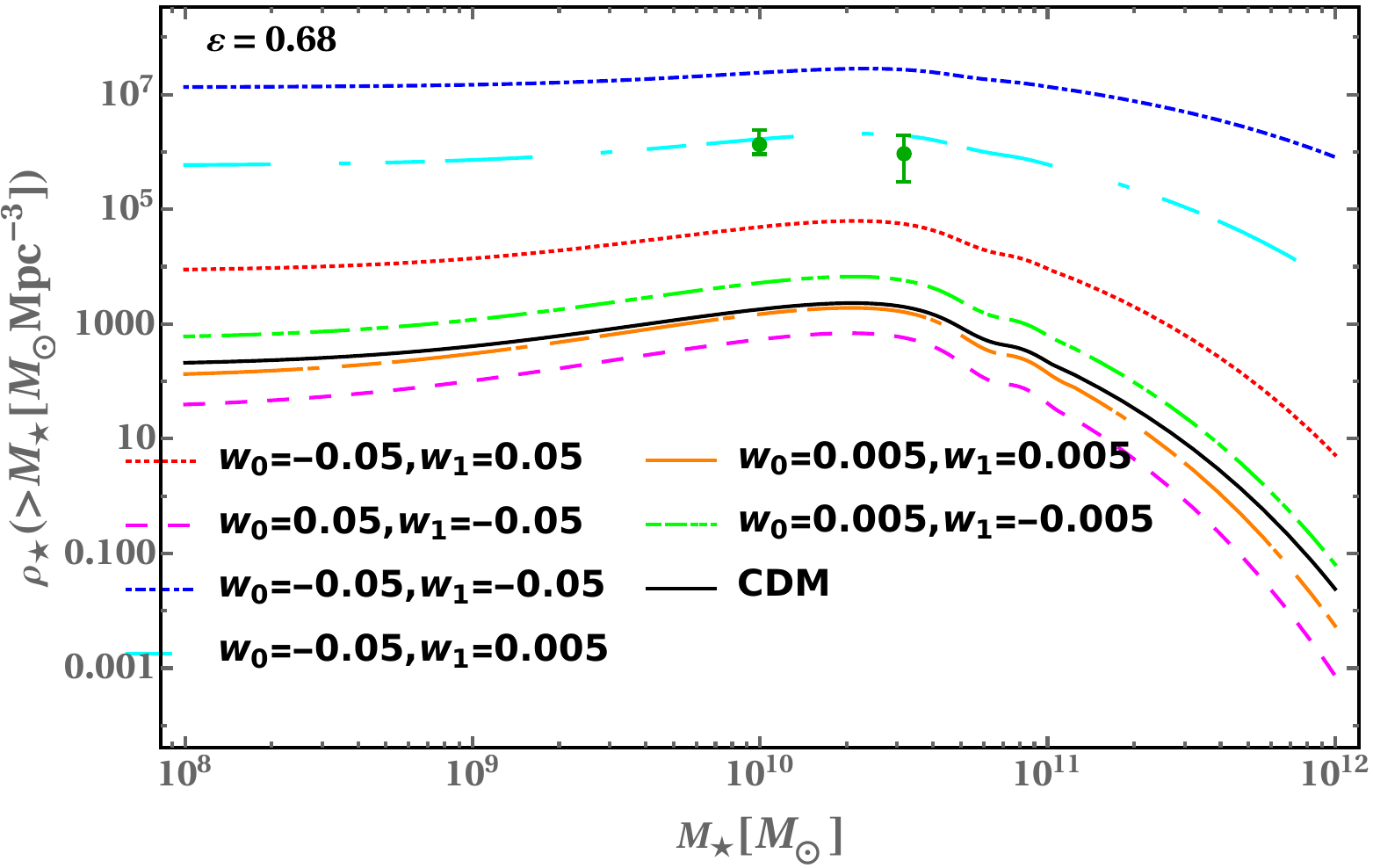}
	\includegraphics[height=4cm,width=5cm]{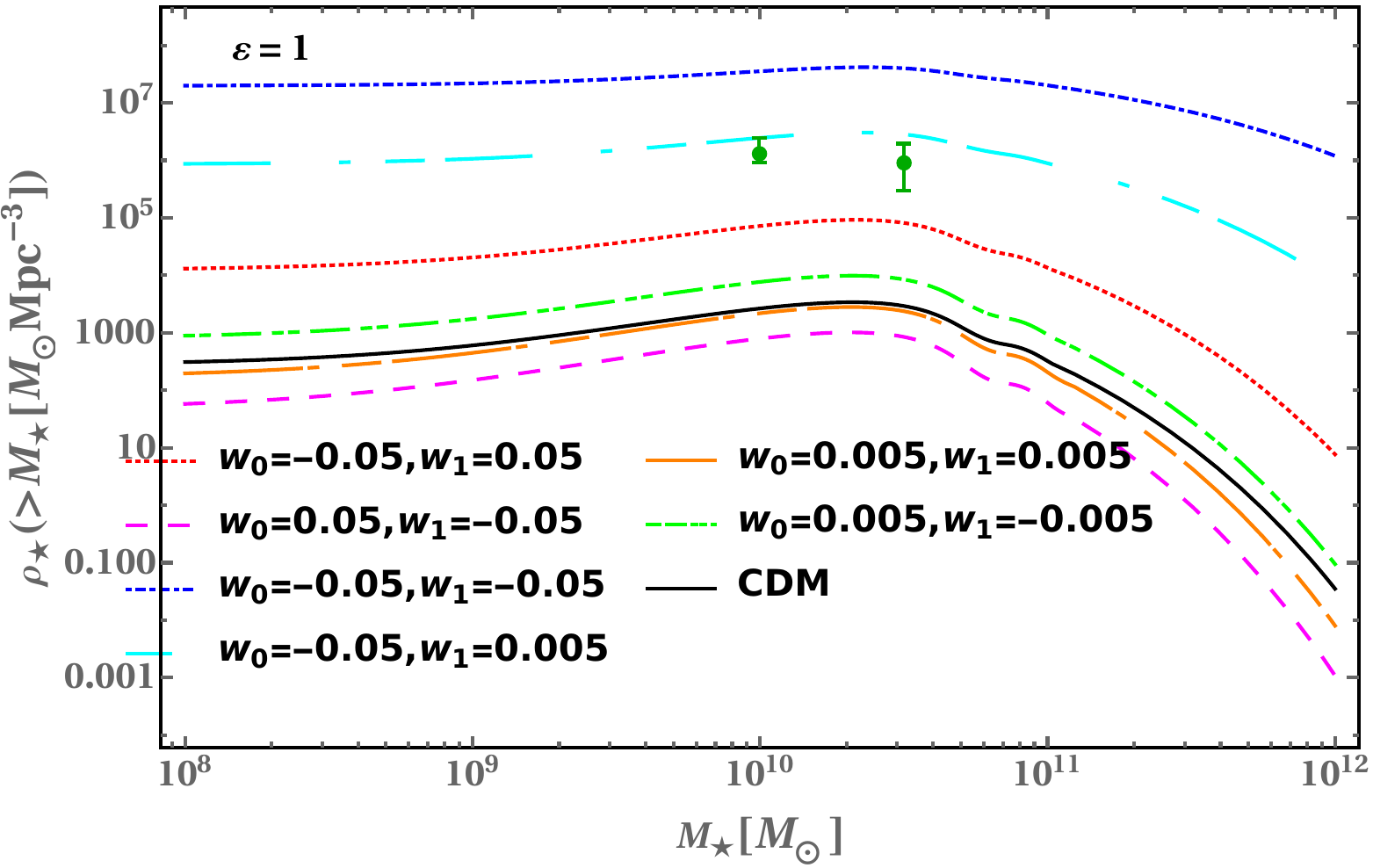}
	\caption{The comoving stellar mass density contained within galaxies more massive than $M_\star$ at $z = 10$ for
		the 10 different dark matter models (4 NDM-cte and 6 NDM-CPL models) for various values of the assumed conversion efficiency $\epsilon = 0.32, 0.68$ and $1$ of a halo's cosmic allotment of baryons into stars.}
	\label{fig7}
\end{center}
\end{figure*}
Here, we calculate the cumulative comoving stellar mass density within galaxies with stellar mass above $\rho_\star(>M_\star)$ for different DM models at three redshifts probed by the analysis of JWST CEERS data: $z = 7.5$,  $z = 9.1$, and $z = 10$. In Figure~\ref{fig6}, we show the theoretical predictions for the comoving cumulative stellar mass density ($\rho_\star(>M_\star)$) in terms of $M_\star$. We assume three different values of the efficiency of converting gas into stars $\epsilon = 0.32, 0.68$, and the conservative value, $\epsilon = 1$ which maximizes the stellar mass predicted by a given scenario. We find that both alternative DM models and CDM are consistent with JWST data at $z = 7.5$ when the assumed conversion efficiency, $\epsilon = 1$. Notably, the NDM-CPL model is compatible with the JWST data for all three values of $\epsilon$ at $z = 7.5$. At $z = 9.1$, the CDM and NDM-cte models fail to explain the JWST data, but NDM-CPL with $\epsilon=1$ manages to increase the  $\rho_\star(>M_\star)$ up to the level claimed by JWST. However, at $z = 10$, all three theoretical models under study predict a lower value compared to the JWST observations.

\begin{figure}
\begin{center}
	\includegraphics[height=4cm,width=8cm]{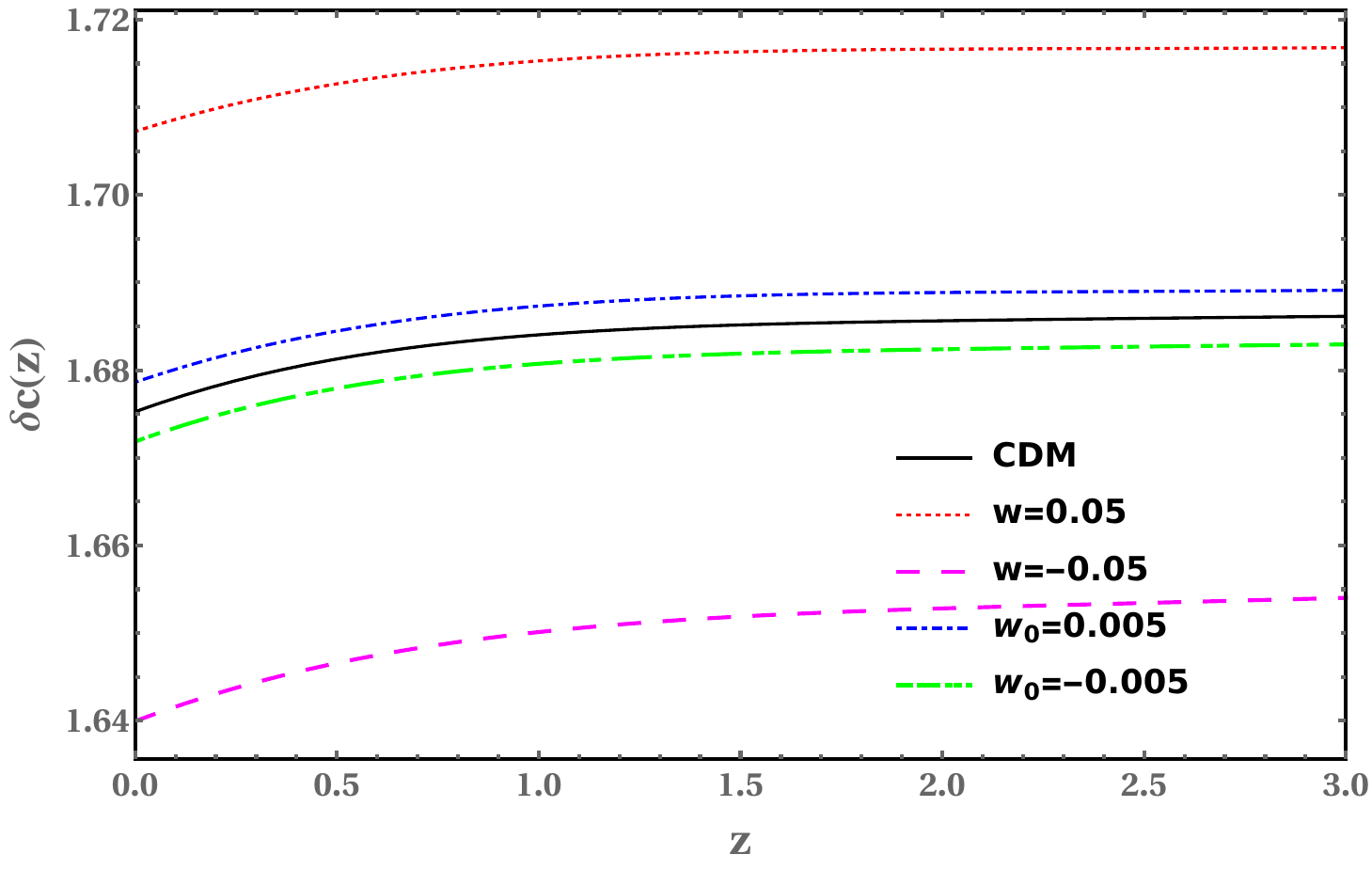}
	\includegraphics[height=4cm,width=8cm]{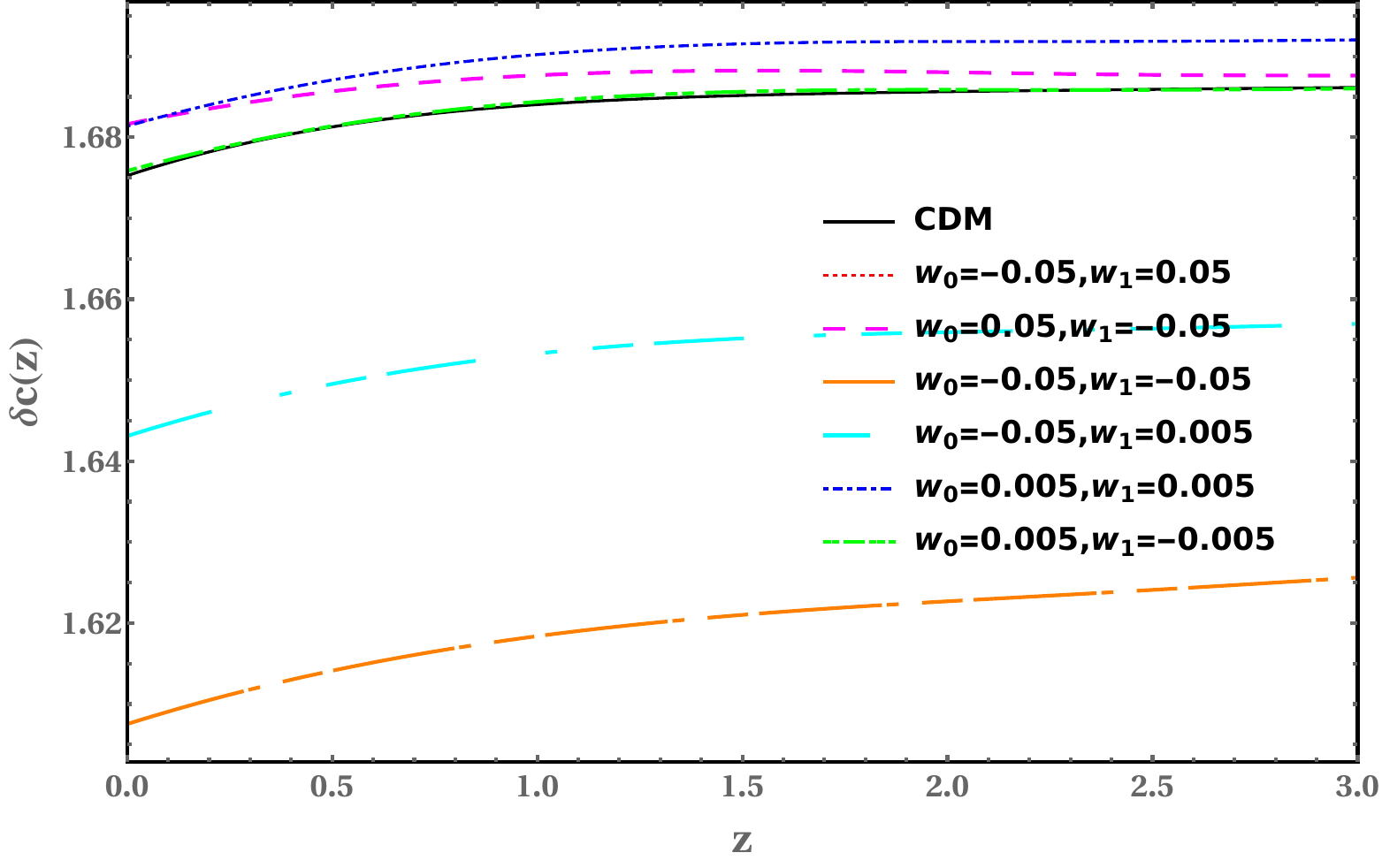}
	\caption{Evolution of the critical collapse density, $\delta_c$, as a function of redshift, for the different dark matter models.}
	\label{fig8}
\end{center}
\end{figure}

\begin{figure}
\begin{center}
	\includegraphics[height=4cm,width=8cm]{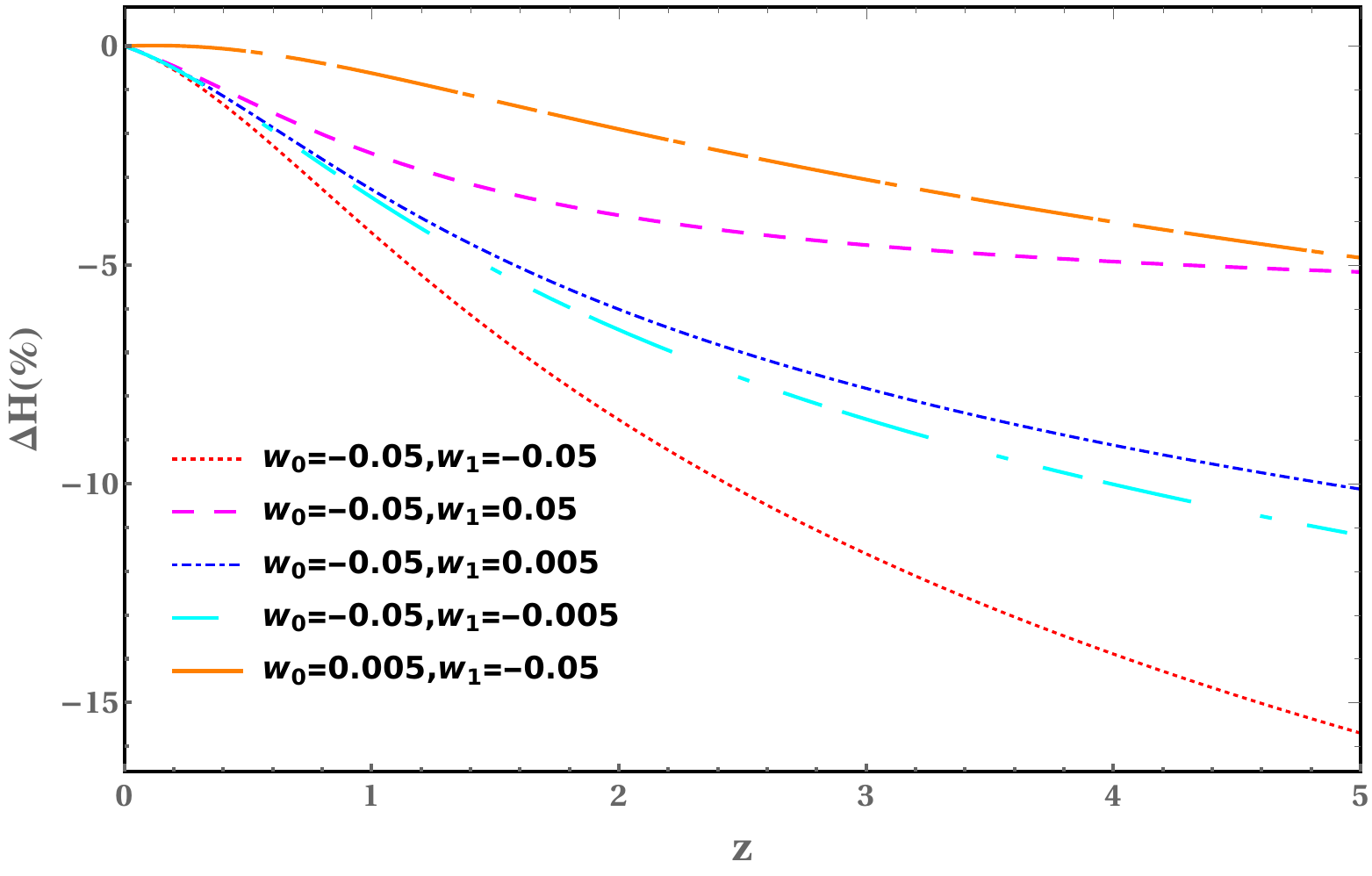}
	\caption{Evolution of relative Hubble parameter, $\Delta H(z)$, as a function of redshift, for the different  negative values of dark mater EOS in NDM-CPL model .}
	\label{fig10}
\end{center}
\end{figure}

So far, we have focused on the best-fit values for the parameters in the different scenarios for the DM equation of state. Next, we plot $\rho_\star(>M_\star)$ for different values of those parameters. Specifically, we consider 10 different models (4 NDM-cte and 6 NDM-CPL models), corresponding to various arbitrary values of the parameters $w_0$ and $w_1$ for both dark matter models. Our goal is to investigate the effect of changing the dark matter equation of state in justifying the high redshift $z = 10$ observed by JWST. For this purpose, we set the values of other free cosmological parameters to be the same in all cases as $\Omega_m= 0.3$ and $H_0 = 70\rm kms^{-1}Mpc^{-1}$. We show the result in Figure~\ref{fig7} and observe an interesting point in these plots: for smaller negative values for both models, we achieve better agreement with the observed values.  Specifically, the theoretical predictions for $\rho_\star(>M_\star)$  in the NCDM-cte model for $w = -0.05$ and in the NDM-CPL model for $w_0 = -0.05$ and $w_1 = 0.005$ (noting that at the high redshifts, the CPL EoS tends to $w_0 + w_1$ ) are consistent with the observational data at $z = 10$ for various values of the assumed conversion efficiency $\epsilon$. In Figure~\ref{fig8}, we show that for these two cases which are in agreement with the JWST observational data, the different values of the critical density obtain smaller values than the standard model. Figure \ref{fig1} indicates that smaller values of $w$ lead to a slower cosmic expansion compared to the standard CDM model. This slower expansion rate increases the comoving stellar mass density within galaxies, as depicted in Figure~\ref{fig7}. The increased stellar mass density could be a direct consequence of the altered dark matter dynamics and its effect on the rate of expansion. We have shown in figure \ref{fig10} how the change of $w_0$ and $w_1$ individually affects the decreasing behavior of the Hubble parameter. By providing this detailed analysis, we aim to clarify why the curves with higher parameter values still show similarities to the standard CDM model, and how the combined effect of  $w_0$ and $w_1$ influences the overall results.

In our computations, we have assumed that the critical overdensity, $\delta_c$, is independent of the mass scale and depends only on redshift. However, since the equation of the state of alternative DM models is nonzero, it is more appropriate to consider its dependence on mass, $\delta_c(z, M)$, we leave for future studies~\citep{Herrera:2017epn, Lin:2023ewc}.
\section{Conclusions}
\label{concl}
In a flat FLRW Universe, we investigated the perturbations of dark matter with different non-zero equations of state in the nonlinear regime using the spherical collapse method. This efficient analytical approach allows us to study structure formation in the nonlinear regime.  Initially, we constrained our models with novel cosmological data, including the luminosity distance of type Ia supernovas, cosmic chronometers, measurements of the growth rate of structure, and the CMB temperature and polarization angular power spectra. Using the Markov Chain Monte Carlo method, we obtained the current values of the free and cosmological parameters necessary for validating the spherical collapse model. Next, we computed the spherical collapse parameters: the linear overdensity $\delta_c(z_c)$, the virial overdensity $\Delta_{{}_{\rm vir}}(z_c)$, and the overdensity at the turnaround $\xi(z_c)$. Our findings indicate that: (i) the linear overdensity $\delta_c$ approaches the expected value of $\delta_c = 1.686$ in the EdS limit at high redshifts, (ii) the virial overdensity $\Delta_{{}_{\rm vir}}$ and the overdensity at the turnaround $\xi$ converge to 178 and 5.55, respectively, at early times, consistent with the EdS Universe, where the impact of dark energy is negligible.  Our results imply that $n(>M)_{\rm mod}/n(>M)_{\rm CDM}$ is lower in the NDM-cte model and higher in the NDM-CPL model compared to the CDM results.

In the final section, we addressed the unexpectedly high stellar mass densities of massive galaxies at $7\lesssim z\lesssim10$ observed by the early JWST, which suggests a higher star formation efficiency and potential tension with the standard CDM model. We examined the comoving stellar mass density within galaxies more massive than $M_\star$. Our results demonstrate that the NDM-CPL model aligns with JWST data for all values of $\epsilon$ up to $z = 9$. However, for higher redshifts ($z = 10$), JWST observations indicate an excess in the comoving stellar mass density beyond the predictions of our theoretical models.

As evident from Figure~\ref{fig7}, having dark matter with a negative equation of state parameters ($w_1<0$) significantly enhances the theoretical predictions for cumulative stellar mass density, thereby improving the agreement with JWST results.


\section{ACKNOWLEDGMENTS}

Z.D. was supported by Iran Science Elites Federation Grant No M401543 and the Korea Institute for Advanced Study Grant No 6G097301. A.A. also received funding support from the European Union's Horizon 2020 research and innovation programme under the Marie Sk\l{}odowska -Curie grant agreement No 860881-HIDDeN.

\section{DATA AVAILABILITY}
No new data were generated or analysed in support of this
research.


\bibliographystyle{mnras}
\bibliography{ref}
\end{document}